\documentclass[preprint,12pt,authoryear]{elsarticle}

\usepackage[english]{babel}
\usepackage[T1]{fontenc}
\usepackage{subfig}
\usepackage{float}
\usepackage{microtype}
\usepackage{lscape}
\usepackage{appendix}
\usepackage{siunitx}
\usepackage[dvipsnames]{xcolor}

\usepackage{amsmath}
\usepackage{graphicx}

\journal{Mathematical Psychology}

\begin{document}

\begin{frontmatter}

\title{Interval Timing: Modelling the break-run-break pattern using start/stop threshold-less drift-diffusion model}

\author[inst1]{Jason Zwicker}
\ead{jason.zwicker@rmc.ca}

\author[inst1]{Francois Rivest\corref{cor1}}
\ead{francois.rivest@\{mail.mcgill.ca, rmc.ca\}}

 \cortext[cor1]{Corresponding Author}

\affiliation[inst1]{organization={Royal Military College of Canada},
            addressline={13 General Crerar Cres}, 
            city={Kingston},
            postcode={K7K 7B4}, 
            state={Ontario},
            country={Canada}}
            
\date{June 2021}

\maketitle

\begin{abstract}

Animal interval timing is often studied through the peak interval (PI) procedure.  In this procedure, the animal is rewarded for the first response after a fixed delay from the stimulus onset, but on some trials, the stimulus remains and no reward is given.  The standard methods and models to analyse the response pattern describe it as break-run-break, a period of low rate response followed by rapid responding, followed by a low rate of response.  The study of the pattern has found correlations between start, stop, and duration of the run period that hold across species and experiments.  

It is commonly assumed that to achieve the statistics with a pacemaker accumulator model, it is necessary to have start and stop thresholds.  In this paper, we will develop a new model that varies response rate in relation to the likelihood of event occurrence, as opposed to a threshold, for changing the response rate.  The new model reproduced the start and stop statistics that have been observed in 14 different PI experiments from 3 different papers.  The developed model is also compared to the two-threshold Time-adaptive Drift-diffusion Model (TDDM), and the latest accumulator model subsuming the scalar expectancy theory (SET) on all 14 datasets.  The results show that it is unnecessary to have explicit start and stop thresholds or an internal equivalent to break-run-break states to reproduce the individual trials statistics, the average behaviour, and the break-run-break analysis results.  The new model also produces more realistic individual trials compared to TDDM.
\end{abstract}

\begin{highlights}
\item Pacemaker-accumulator model without start and stop thresholds.
\item Analysis of the peak-interval procedure including: individual trial, start and stop, and the average response curve.
\item Linearly increasing response rate until expected time of reward.
\item Drift-diffusion model with a probabilistic response for the peak-interval procedure.
\end{highlights}

\begin{keyword}
interval timing \sep peak-interval \sep drift-diffusion model \sep threshold
\end{keyword}

\end{frontmatter}
\newpage
\section{Introduction}\label{intro}

The ability to perceive the passage of time, from seconds to minutes, is defined as interval timing.  Timing the duration of an interval is fundamental to a wide range of processes, including decision making, planning, reaction to a predator, reaction to environmental conditions, or timing a repeated event \citep{gallistel1990organization}. Interval timing has been studied for decades, yet there is still debate on some of the properties and behaviours that characterize timing.  In the study of interval timing, a rich set of procedures and models have been developed that shape how we think about animal interval timing.

The Peak Interval (PI) procedure \citep{roberts1981isolation,catania1970reinforcement} is a variation of the fixed interval (FI) procedure used to examine interval timing.  In the FI procedure, a subject is exposed to a stimulus and the first response after a fixed interval of time from the stimulus onset is rewarded.  The PI procedure introduces probe trials where the stimulus stays on longer and no reward is received.  After several probes, animals learn to give up if they do not get the reward at the expected time \citep{balci2009acquisition}. The analysis of the resulting response pattern has influenced many models and theories about interval timing.  In particular, a response pattern of break-run-break has been observed \citep{church1994application}.  In the analysis and modeling of this pattern, pacemaker-accumulator (PA) models have been shown to require two thresholds \citep{gibbon1990representation,church1994application,luzardo2017drift} determining the start and stop of the animal response pattern (the run).

There is still debate about the subject's response to the peak procedure at the individual trial level \citep{staddon2003operant}. Does the subject wait, respond at a constant rate, then wait again \citep{church1994application}?  Are there more than two distinct identifiable rates of response such as is suggested by the higher rate in the run period identified in \citet{cheng1993analysis}; or perhaps simply an increasing rate as event probability increases as suggested by the increasing frequency in Packet Theory \citep{kirkpatrick2002packet, kirkpatrick2003tracking}?  Is there an observable pattern of response before and after the period of high rate responding?  These questions can be summarized as the following: Does the response rate continually change, or are there constant low and high response rates (break-run-break)?  There is evidence for both conclusions \citep{church1994application,kirkpatrick2002packet, kirkpatrick2003tracking}.

Many of the properties of interval timing have been successfully modeled based on break-run-break data.  This has included a significant amount of analysis of the start and stop of the run period.  The descriptive statistics of the start and stop of the high rate of response are well understood and consistent across studies and data sets.  The analysis of start and stop times across trials revealed \citep{church1994application, gallistel2004sources, luzardo2017drift}:

\begin{itemize}
\item Positive correlation between start and stop
\item Negative correlation between start and duration
\item Positive correlation between duration and middle
\item Coefficient of variation for the start larger than the stop    
\end{itemize}

There are numerous interval timing models. Many of them focused on the pacemaker-accumulator idea such as the Behavioral Theory of Timing (BeT) \citep{killeen1988}, the Scalar Expectancy Theory (SET) \citep{gibbon1977scalar, church1994application}, and more recently the Time-adaptive Drift-diffusion Model (TDDM) and TopDDM models \citep{simen2011model, rivest2011, simen2013timescale, balci2016decision}.  Packet theory \citep{kirkpatrick2002packet,kirkpatrick2003tracking} is based on bouts of response that increase in likelihood as the likelihood of reward increases.  Other approaches are based on neural oscillation, such as Striatal Beat Frequency (SBF) \citep{matell2000neuropsychological,buhusi2005makes,oprisan2011modeling,oprisan2014phase} and a more recent model based on Neural Oscillation States \citep{varella2019model}. Other adaptive models include recurrent neural network approaches \citep{hardy2018encoding, raphan2019modeling} and behavioral Learning to Time Theory (LeT) \citep{machado1997learning, pinheiro2016learning}. Finally, in \citet{hasegawa2015model}, they developed a multi-clock model based on LeT and SET that provided similar results to SET. 

SET has been compared to BeT, LeT, and Packet Theory \citep{church1994application, yi2007applications, lejeune2006scalar}. Based on these comparisons, SET provides the most accurate model of interval timing.  In addition, BeT and LeT have been directly questioned in \citet{lejeune2006scalar}, and later Machado revised LeT by adding SET's timing mechanism \citep{machado2009learning}.  Most interval timing models focused solely on reproducing the average response curves without looking at individual trial statistics such as the start and stop time and their correlations. The only exception we found are SET, TDDM, and the Neural Oscillation States model \citep{church1994application, luzardo2017drift,varella2019model}. 

In particular, TDDM is an accumulator-based break-run-break model that was shown to have all SET's important properties \citep{luzardo2017drift} while also being able to reproduce the learning dynamic of animal interval timing \citep{luzardo2013adaptive, luzardo2017drift}. It also has an equivalent spiking neuron model \citep{simen2011model}. The single-threshold FI model is often called TopDDM. TDDM was shown to be highly accurate across an extensive range of data sets.  In this paper, by TDDDM, we explicitly refer to the two-threshold TDDM model for PI data from \citet{luzardo2017drift} which will be the standard for comparison. 

Both PA threshold-based models, SET and TDDM, concluded that two thresholds and a single accumulator were required to model break-run-break statistics of the PI data \citep{church1994application, luzardo2017drift}; the thresholds being used to represent the start and stop decision boundaries. Nevertheless, the question remains, is the break-run-break assumption required to explain or reproduce the animal behaviours?

In order to answer if the peak interval response pattern is strictly break-run-break or if the response rate increases as the likelihood of reward increases, we developed a new model, called the Probabilistic Response Drift-diffusion Model (PRDDM), whose response rate is linearly dependent on elapsed time. This model, while using the same accumulator as TDDM, does not by definition have break-run-break states. Instead, its probability of response continuously changes  over time in a way similar to packet theory \citep{kirkpatrick2002packet,kirkpatrick2003tracking}. We compare this new model to TDDM on individual trials, start and stop statistics, and all of the subjects behaviours, showing that it reproduces the animal data equally well or better. Furthermore, the results indicate that response rates increase with the likelihood of reward is sufficient to reproduce the data, including the break-run-break statistics. This suggests that having start and stop thresholds (or distinct break and run states) is not required and may only be a consequence of our assumptions. To our knowledge, this is the first PA model reproducing those properties not requiring this assumption.

\section{Models}

The Probabilistic Response Drift-diffusion Model (PRDDM) has a response rate that varies proportionally with the distance in time from the subject's estimated time of reward.  While the TDDM accumulator is maintained, the thresholds that determine the start and stop of the response period are removed. This gives a simpler model that also generates more realistic trials. We will first describe the basis of drift-diffusion models, TDDM, and finally the new PRDDM.

\subsection{Drift-diffusion Model (DDM)}

Drift-diffusion models \citep{ratcliff1978theory} have been used to model a wide range of real-time decision making paradigms \citep{voss2013diffusion} such as two alternative forced action tasks \citep{ratcliff2008diffusion}, and interval timing \citep{luzardo2017drift}.  The model value is estimated by accumulating noisy evidence until a decision boundary is reached.  The computational representation is a stochastic differential equation that can be incrementally calculated.

A DDM has three parameters: drift, diffusion, and decision boundary.  The drift is the rate at which the accumulator approaches the decision boundary, while diffusion is the noise applied to the drift at each time step.  Finally, the decision boundary represents a decision or internal change of state.

The stochastic differential equation for a drift-diffusion model is
\begin{equation} \label{sto_dif_eqn}
\mathrm{d}x = A \, \mathrm{d}t + m\, \mathrm{d}W,
\end{equation}
where $x$ is the particle position, $A$ > 0 is the model drift rate, $m$ > 0 is the scale of diffusion, and $\mathrm{d}W$ is the noise in the system.  This can be incrementally approximated by
\begin{equation} \label{inc_ddm}
x(t+\Delta t) = x(t) + A \Delta t + m \mathcal{N}(0,\,1)\  \sqrt[]{\Delta t}, \end{equation}
where $\Delta t$ is the change in t, or the time-step, and $\mathcal{N}(0,\,1)$ is a random variable from the normal distribution, with a mean of zero and a standard deviation of one, that provides the random component of the noise term.

In a DDM, a decision is reached when $x(t) > z$, where $z$ is the decision threshold.  The threshold $z$, may be a reflecting boundary, an absorbing boundary, or just a decision threshold. If reflecting, it continues to accumulate with a negative drift; if absorbing, the value of $x(t)$ remains $z$ after the decision is made; and if just a decision threshold, it will continue to accumulate evidence beyond the value of $z$.

\subsection{Time-adaptive Drift-diffusion Model (TDDM)}
\label{CH_PI_TDDM_Parameters}

When modeling the peak-interval procedure, TDDM \citep{luzardo2017drift}, like SET \citep{church1994application}, is a two-threshold break-run-break model of response. TDDM uses a drift-diffusion accumulator ($x$), a noisy start threshold ($z_1$), and a stop threshold ($z_2$) where $0 < z_1 < 1 < z_2$.  The model starts to respond when $x > z_1$,  at which time it responds with the high rate until $x > z_2$, at which time it returns to the low rate. The estimated time of reward is considered to be when $x(t) = 1$.  

TDDM has five free parameters: drift, diffusion, start threshold, stop threshold, and standard deviation of the start threshold. All parameters, except for drift, are calculated from the start and stop statistics of the experimental data.  The drift rate can be estimated from the FI, or learned automatically to model animal learning dynamics \citep{luzardo2013adaptive}. For further details, see \citet{luzardo2017drift}.  

As in SET, the two thresholds allow the model to flip between two states: break and run, naturally generating break-run-break patterns. While TDDM simulates single trials, the simulations only provide constant responding during the high rate period.  

\subsection{Probabilistic Drift-diffusion Model (PRDDM)}

PRDDM has two processes:  a drift-diffusion process and a stochastic response.  The drift-diffusion process gathers evidence of elapsed time.  We will interpret the accumulator's value as the subject's estimate of the likelihood of reward occurrence at time $t$.  The probability of response to the stimulus will be modeled as a function of this likelihood. 

PRDDM has a single reflecting boundary that does not represent the response period's beginning or end.  The value of the accumulator is strictly interpreted as the likelihood of event occurrence (likelihood of reward occurring).  The time when the accumulator equals, or first exceeds, the boundary ($z=1$) is interpreted as the estimated time of reward.

The probability of response is defined by the subject's interest, represented by the peak mean response rate, in the reward and the estimated likelihood of reward.  In this model, the likelihood of reward is the subject's estimate of the likelihood of the reward occurring at time $t$, based on the estimated elapsed time.  The likelihood of response is modelled as a scalar constant ($\tau$, subject interest in the reward) multiplied by the likelihood of event occurrence ($x(t)$, current accumulator value).  The model is entirely defined by Eq.~\ref{inc_ddm}, the reflecting boundary ($z=1$), and the probability of response ($R$) given by
\begin{equation} 
\label{ProbResp}
\mathrm{P}(R=1 \mid t) = \tau x(t),
\end{equation}
where $0 < \tau < 1$ and $t > 0$ is the elapsed time since the stimulus onset.

\subsubsection{Model Parameters}
\label{Ch_PI_Model_Parameters}

\begin{figure} [!ht]
\centering

\begin{tabular}{|c | c|}
\hline
\subfloat[PRDDM Process]
{
\includegraphics[width=0.45\textwidth]{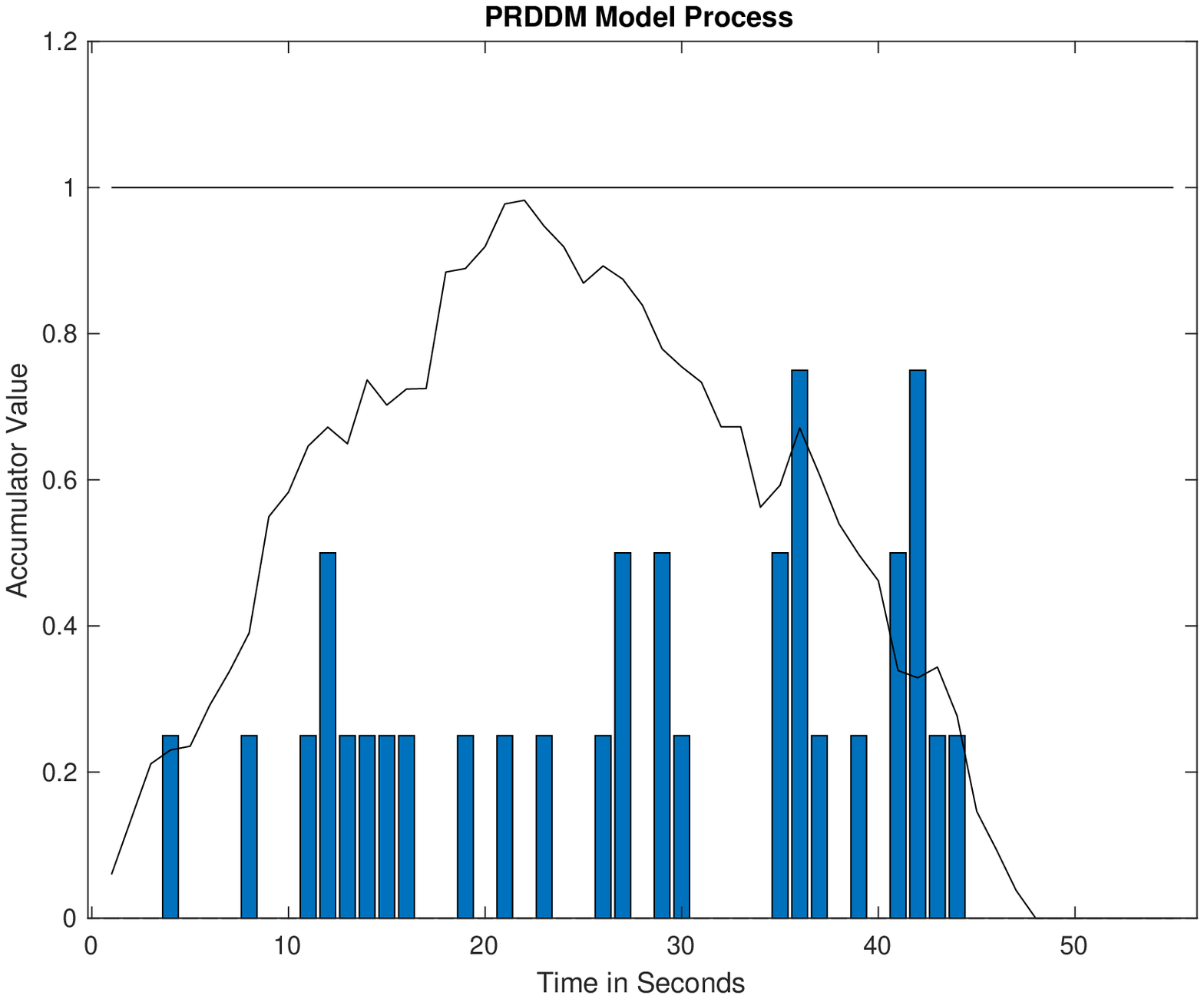}}

&\subfloat[TDDM Process]
{
\includegraphics[width=0.45\textwidth]{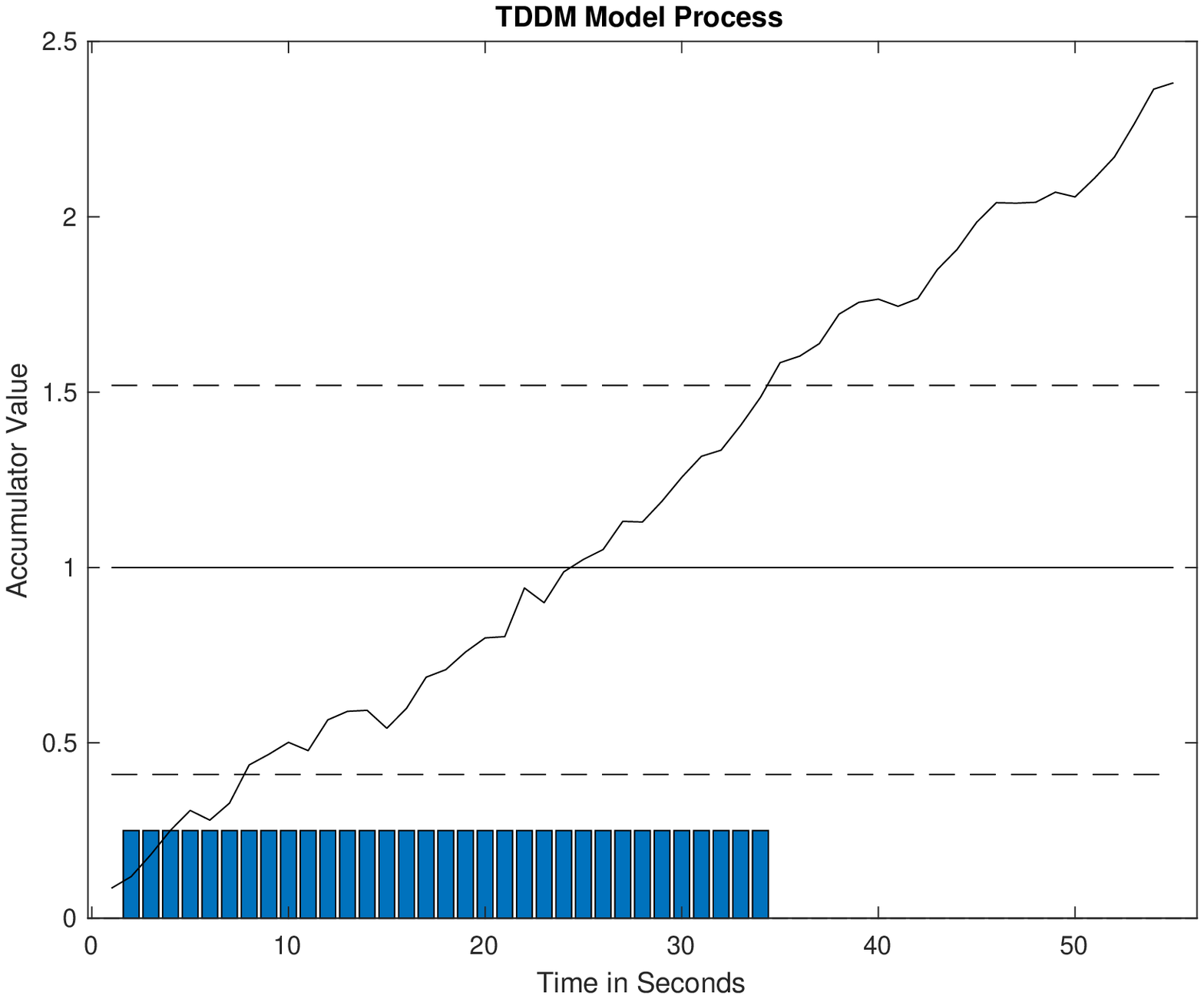}}

\\
\hline

\end{tabular}

\caption[TDDM and PRDDM Process]{\label{Fig_Model_Process} An example of (a) PRDDM's and (b) TDDM's process including accumulator (solid line), start and stop thresholds (dashed horizontal lines) and an example single trial response (blue histogram).}

\end{figure}

In PRDDM, the DDM is the pacemaker-accumulator, and it models the subject's internal clock and  accumulator.  Drift is the rate of evidence accumulation, diffusion is the noise in the accumulation, and the boundary is the subject's estimated time of reward.  See Fig.~\ref{Fig_Model_Process}(a) for an example of the DDM accumulator.  PRDDM has one more parameter: the mean response rate, which controls the subject's rate of response.  Thus, PRDDM has four parameters: three calculated (drift, diffusion, and mean response rate) and one fixed parameter (reflecting decision boundary). 

Drift, $A$, is the rate at which the time estimation process gathers evidence and estimates time remaining; see Eq. \ref{inc_ddm}. In PRDDM, we estimate the value of $A$ as one divided by the expected value of the peak time from the data to be modeled.  Mathematically, 
\begin{equation} \label{param_calc_A_eqn}
A = \frac{1}{\mathrm{E}[pt]},
\end{equation}
where $pt$ is the peak time of a trial and $\mathrm{E}[\cdot]$ is the expected value operator. 

Diffusion, $m$, is the amount of noise in the time estimation process.  It is applied at every incremental update; see Eq.~\ref{inc_ddm}.  In a DDM, the points in time when the accumulated value first passes the boundary (first passage) forms an inverse Gaussian distribution. Since PRDDM estimates the peak time as the first passage of the decision boundary ($z$), the value of $m$ can be estimated from the variance of the peak time estimated for each trial.

An inverse Gaussian distribution is denoted by $IG(\mu,\lambda)$, where $\mu$ is the mean and $\lambda$ is the shape parameter.  When considering a DDM and first passage time of the decision boundary, $\mu$ and $\lambda$ are approximately $\mu = 1/A$ and $\lambda = 1/m^2$.  As the variance of an inverse Gaussian variable $X$ is $\text{var}(X) =  \mu ^ 3 / \lambda$, we can then substitute the values for $\mu$ and $\lambda$ into the equation for the variance of an inverse Gaussian distribution.  This yields 
$\text{var}(X) = m^2 / A^3$.  Since peak time is estimated by the first passage time, $\text{var}(pt) = \text{var}(X)$.  This can be transformed into $m^2 = A^3 \, \text{var}(pt)$ and used to estimate the value of $m$:
\begin{equation} \label{param_calc_m_eqn}
m = \sqrt{A^3 \, \text{var}(pt)}.
\end{equation}

Peak mean response rate, $\tau$, is a calculated value. It is twice the mean of the mean response rate of the individual trials.  The mean response rate for each trial is calculated in the interval $[0, 2pt]$, as the model simulates that duration.  As PRDDM is a linear approximation, the mean response rate for the interval is half of the response rate at the peak.  Thus the mean response rate is multiplied by 2. This resulting value is the mean response rate at peak time from the data to be modelled.

Reflecting boundary, $z$, in conjunction with the drift,  controls the subject's estimated time of reward.  As it is only related to the drift, without loss of generality, the boundary can be fixed at $z=1$.

\subsubsection{PRDDM compared to TDDM}

An example of PRDDM and TDDM is shown in Fig.~\ref{Fig_Model_Process} (a) and (b) respectively. While both utilize DDMs for the pacemaker-accumulator, the decision to respond and response rates differ greatly. PRDDM utilizes the accumulator value to estimate the likelihood of reward occurrence at each time step, and this likelihood then determines the model's probability of response. Therefore, PRDMM will likely show more responses during the middle section of the accumulation process than at both ends of a probe trial. However, since it is probabilistic, any small time window may have a different response rate, generating more animal-like trials. In contrast, TDDM shows a clear break-run-break pattern with fixed low and high response rates. Although PRDDM is also a pacemaker-accumulator model, in contrast with SET and TDDM, it does not assume two-decision thresholds. Instead, the probability of responding is proportional to the accumulator content value, and it assumes a single decision point representing the point where the perceived elapsed time matches the learned time interval. This is a major difference with SET \citep{church1994application} which explicitly generates start and stop time points for the break-run-break pattern. By having a linear response rate, PRDDM has no built-in start or stop inflection (or decision) points whatsoever. Nevertheless, as we will show in the next sections, when PRDDM results are analyzed using the break-run-break assumption, it reproduces the start and stop statistics as well as break-run-break models. Thus, showing that the underlying animal process does not need to include an explicit sequence of break, run and break states, unlike SET's assumption.

\section{Methods}

The objectives of this experiment are the following: validate PRDDM against existing data; show that start and stop thresholds are not required in PA models; and that the break-run-break pattern of responding can be achieved without any underlying explicit break-run-break process.

In order to achieve this, PRDDM and TDDM will be used to simulate 16 PI datasets:  three from \citet{church1998temporal}, 12 from \cite{kirkpatrick2000independent}, and one from \citet{balci2009acquisition}.  The results from each simulation will be compared to the statistics of the dataset that it is simulating.  To support this comparison, the models will produce the same number of trials and trial duration as the original dataset, with all model parameters for both models estimated from the animal data using the given formulae.  PRDDM parameters will be estimated as described in section~\ref{Ch_PI_Model_Parameters}. TDDM (which also represents SET) parameters will be estimated as in \citet{luzardo2017drift}. 

Data analysis for both simulations and animal data will be the same throughout. In supporting the comparative analysis between the simulation and experimental datasets, the simulation results will be grouped into three sections as follows: individual trial, start and stop analysis, and average behaviour, as detailed below:

\begin{itemize}

\item \textbf{Individual trials} will be used to compare mean response rate, mean number of responses, calculated peak time, and the individual trial response pattern.  
    
\item \textbf{Start and Stop} will be compared by calculating the mean absolute error of the start, stop, mid-point, duration, coefficient of variation (CV), and of the mean of the correlations between start to stop, start to duration, and mid-point to duration for each model.  Finally a qualitative comparison of the dataset to the simulation based on graphing the alignment of the start and stop times will be performed.

\item \textbf{The average} behaviour will be compared by the Akaike Information Criteria (AIC) with small size correction ($AIC_c$) of the average response curve between each model and the experimental data set.
\end{itemize}

Details of the data set and data analysis are provided in the following two subsections, respectively.

\subsection{Datasets}

\subsubsection{Balci 2009}

In 2009, Balci published a study that investigated the acquisition of timed responding in the peak interval procedure by mice \citep{balci2009acquisition}. This data set is from the raw data collected during the experiments supporting the 2009 paper. 

Forty-five experimentally na\"{i}ve mice were trained to press a lever on cue, responding to a 10 second FI, and finally a 30 second FI.  Once the mice successfully completed the pre-training, the PI procedure was introduced.  There was a total of 16 sessions lasting approximately 80 minutes each.  Each session consisted of 30 FI trials and 15 PI trials randomly interspersed.  The PI trials were 90 seconds in length (3 times the FI period), and the inter-trial interval (ITI) was uniformly distributed between 10 and 30 seconds.  We used steady-state data from sessions 15 and 16 \citep{balci2009acquisition}. 

\subsubsection{Church 1998}

In 1998, Church et al. published a study that included an experiment that examined the peak procedure using rats with fixed and uniform intervals \citep{church1998temporal}. The experiment had fixed intervals of 30, 45, and 60 seconds.  The uniform interval was from 30-60 seconds.  From these trials, we extracted three datasets: PI trials for 30 seconds (Church30), 45 seconds (Church45), and 60 seconds (Church60).  

The 20 subjects were first exposed to 60 second fixed time schedule.  The following two sessions were continuous reinforcement with each lever press delivering reinforcement.  For testing, the rats were divided into four group of five rats each.  Each group was exposed to either an FI of 30 second, 45 seconds, 60 seconds, or to the uniform interval.  The peak procedure consisted of a 50 sessions of a random mixture of reinforced and non-reinforced trials.  The non-reinforced trials lasted for 240 seconds plus a mean random interval of 60 seconds.  The inter-trial intervals were 20 seconds plus a mean random interval of 10 seconds.  The first five minutes of each two session included only reward trials.  The remainder of the session included reinforced and non-reinforced sessions at an equal likelihood.  As in \citet{church1998temporal}, we used the data from sessions 21 to 50 for analysis.

\subsubsection{Church 2000}

In 2000, Kirkpatrick et al. published a study that included an experiment using rats that examined stimulus duration and cycle length (duration between rewards) \citep{kirkpatrick2000independent}.  The experiment used FIs of 15, 30, 60, and 120 seconds, with 90, 180 and 360 second cycles.  All FIs were used with a cycle of 180, while 60 also used 90 and 360.  In addition, the stimulus was delivered as either white noise or light.  The experiment included both omission testing and the PI procedure.  From this dataset, we extracted PI datasets with a cycle of 180: 15, 30, 60, 120.  We also extracted an additional four PI 60 datasets, 2 with a cycle length of 90 and 2 with a cycle length of 360.  We label each dataset by `Church 2000 PI Duration Cycle Duration Stimulus', IE. `Church 2000 PI60 180 Light'.  Note that the cycle duration is only included for PI 60 trials.

In that experiment, 48 subjects were divided into two sets of 24, with those sets being sub-divided into groups of 8.  Each group was identified by FI/Cycle (Condition), IE. 60/180.  The subjects in each condition were divided into two, where each subgroup received a different stimulus, either house light or white noise.  When not receiving the stimulus, each subgroup would receive the other signal, IE. If they received white noise as the stimulus, the non-stimulus period would be house light.  The cycle, stimulus and non-stimulus conditions were constant for each subject.  Each training session lasted 105 minutes with a total of 40 sessions being delivered.  The first 40 sessions of training were followed by 12 sessions of trace testing, followed by 13 sessions of the original training.  Then 34 sessions of the PI procedure were executed.  The final trials were omissions and five additional PI trials to return the subjects to the PI conditions.
During the PI procedure, in 25\% of the trials the subject would receive the stimulus for four times the cycle length and would not receive a reward.  A new cycle would begin at the end of the probe trial.  We used the 34 sessions of PI data that were reported in \citet{kirkpatrick2000independent}. 

\subsection{Data Handling and Calculations}

Data from the original experiments and the simulations of TDDM and PRDDM will be handled with identical methods.  This includes binning (using 1 sec bins), start and stop analysis over the interval $[0,2.5FI]$, and normalization of all trials. The calculations and methods are detailed below.

\textbf{Individual Trial Peak Time} is calculated by an iterative method based on the median response from \citet{roberts1981isolation}.  On the first iteration, the median response in the interval $[0,2.5FI]$ is selected.  For example, if there are five responses in the interval, the bin of the third response will be selected as the median response.  On the following iterations, the median of the interval from previous Median-$FI$ to previous Median+$FI$ is selected as the new median.  This process is repeated until the resulting median is within one second of the previous median.

\textbf{Trial Selection} was done according to the start and stop times in comparison with the $FI$ time.  Many effects can impact the determination of start and stop, such as resurgence \citep{kirkpatrick1996cyclic}. These effects can be significant enough to cause start ($S_1$) and stop ($S_2$) to be miscalculated.  Therefore, any trial where $S_1 > FI$, or $S_2 < FI$ are discarded and not included in our analysis.  This is the same method as used in \citet{church1994application} and \citet{gallistel2004sources}.

\subsubsection{Individual Trial}

\textbf{Mean Peak Time} is the mean of the individual trial's peak time. This value is used in the individual trial comparisons.

\textbf{Mean Number of Responses} is calculated over all trials.  First, it sums the number of responses per individual trial for 2.5FI, then takes the mean of that value over all trials.  This value is used in the individual trial comparisons.

\textbf{Mean Response Rate} is calculated over all trials for 2.5FI seconds.  First, the mean number of responses per bin is calculated for each trial. Then, the mean of the mean number of responses is calculated as the mean response rate for all trials.  Finally, this value is used in the individual trial comparisons.

\subsubsection{Start and Stop Analysis}
\label{Sect_StartStop_Calc}
Start ($S_1$) and stop ($S_2$) times of the high rate of response will be determined by using the iterative method from \citet{church1994application} that tests all combinations of $S_1$ and $S_2$, maximizing:
\begin{equation} \label{s1_s2_calc}
\max_{S_1,S_2} \{ S_1 (r- r_1) - (S_2-S_1) (r-r_2) + (ET - S_2) (r - r_3) \}
\end{equation}
for each trial, where $ET$ is the time where the trial ends.  The trial will then be divided into three sections:
\begin{itemize}
\item $t_1 = [0,S_1)$;
\item $t_2 = [S_1,S_2]$; and
\item $t_3 = (S_2,ET]$.
\end{itemize}
such that the response rate for the trial and each section will be calculated:

\begin{itemize}
\item $r$ = response rate of the trial; 
\item $r_1$ = response rate during $t_1$; 
\item $r_2$ = response rate during $t_2$; and
\item $r_3$ = response rate during $t_3$.
\end{itemize}
Once $S_1$ and $S_2$ have been determined, midpoint and duration for each trial will also be calculated:

\begin{itemize}
\item midpoint = $M = \frac{S_1+S_2}{2}$
\item duration = $D = S_2-S_1$
\end{itemize}

\subsubsection{Response Curves}

Since TDDM and PRDDM do not simulate subjects, but instead simulate individual trials, normalization is applied at the trial level before generating response curves. The value of each bin is divided by the maximum value across all trial bins so that maximal response rate per bin is 1.  The response curves are generated from all trials of the simulation or experiment.  \\

\textbf{Akaike's Information Criteria} provides a measure of the distance between a model and a target dataset \citep{burnham1998model}.  It can be used to select the strongest model from a group of models.  AIC was used in \citet{hasegawa2015model} for model comparison.  AIC can be thought of as a measure of likelihood plus a penalty for model complexity and is defined as
\begin{equation} \label{aic_calc}
AIC = -2 \ln (L (\theta)) + 2k,
\end{equation}
where $L(\theta)$, is the log likelihood of the selected model ($\theta$) and $k$ is the number of parameters for the model.  It is suggested that when the sample size ($n$) divided by the number of model parameters is less than 40 that an additional corrective term is used \citep{burnham1998model}.  $AIC_c$ is defined as
\begin{equation} \label{aicc_calc}
AIC_c = -2 \ln (L (\theta)) + 2k + \frac{2k^2 + 2k}{n - k - 1}.
\end{equation}
If all models being compared assume normally distributed errors with constant variance, then AIC can be computed as
\begin{equation} \label{aicc_ls_calc}
AIC_c = n \ln (\sigma^2) + 2k + \frac{2k^2 + 2k}{n - k - 1}
\end{equation}
and
\begin{equation}
\sigma^2 = \frac{1}{n} \sum_{i=1}^n (x_i-y_i)^2,
\end{equation}
where $x$ is the modelled distribution, $y$ is the target distribution, and $n$ is the number of samples from the distributions.  In this paper, $x$ and $y$ are the response distributions of the experimental dataset and the simulated datasets at 1s bin. \\

\textbf{Intra-Class Coefficient} provides a method for determining the distance between two binned datasets.  ICC was used in \citet{swearingen2010pattern}.  In this work, we use the ICC defined by \citet{fisher1955statistical}.  The following equations define ICC:
\begin{equation} \label{icc_calc}
ICC = \frac{1}{n s^2} \sum_{i=1}^{n}(x_i-\bar{x})(y_i-\bar{x}),
\end{equation}
\begin{equation} \label{icc_xbar}
\bar{x} = \frac{1}{2n} \sum_{i=1}^{n}(x_i + y_i),
\end{equation}
and
\begin{equation} \label{icc_ssquared}
s^2 = \frac{1}{2n} \left[ \sum_{i=1}^{n}(x_i - \bar{x})^2 + \sum_{i=1}^{n}(y_i - \bar{x})^2 \right]
\end{equation}
where $x_i$ is sampled from the modeled distribution, $y_i$ is sampled from the target distribution, and $n$ is the number of samples from the distributions as before.

ICC returns a maximum value of one.  The degree to which the distributions are similar increases until the maximum value of one.  \citet{koo2016guideline} suggests the following qualitative categorization of ICC values:
\begin{itemize}
    \item below 0.50: poor;
    \item between 0.50 and 0.75: moderate;
    \item between 0.75 and 0.90: good; and
    \item above 0.90: excellent.
\end{itemize}

\section{Results}

In this section, PRDDM, and TDDM are qualitatively and quantitatively assessed in three areas: individual trial, start and stop statistics and average response curves.  The assessment is presented in the order above.  This section is concluded with a summary of the findings.

\subsection{Individual Trials}

\begin{figure} [t]
\centering

\begin{tabular}{|c|c|c|}
\hline

\subfloat[Church30] {\label{Fig_Church30_Ind_Trial}
\includegraphics[width=0.29\textwidth]{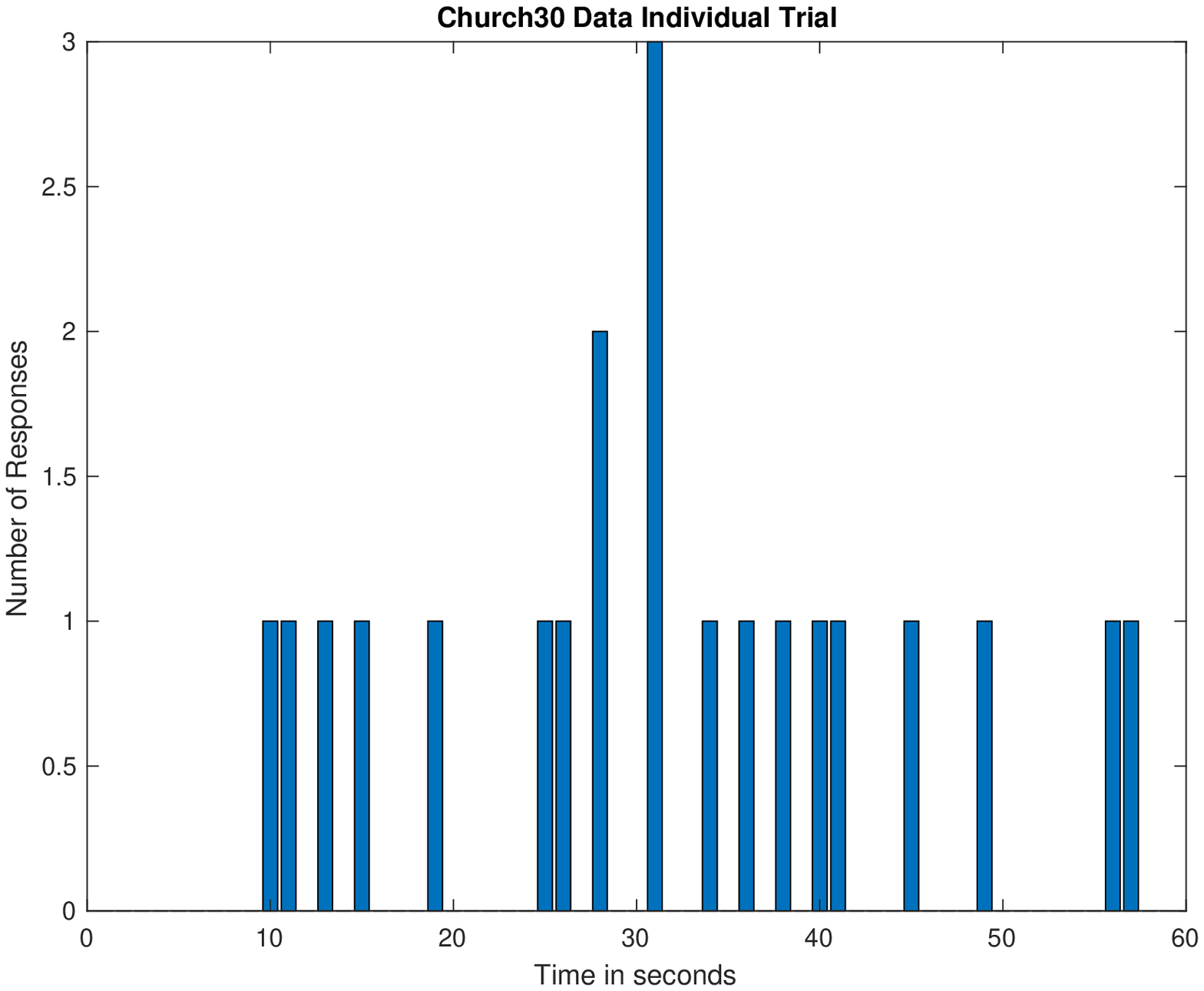}}

&\subfloat[PRDDM] {\label{Fig_Church30_Ind_Trial_PRDDM}
\includegraphics[width=0.29\textwidth]{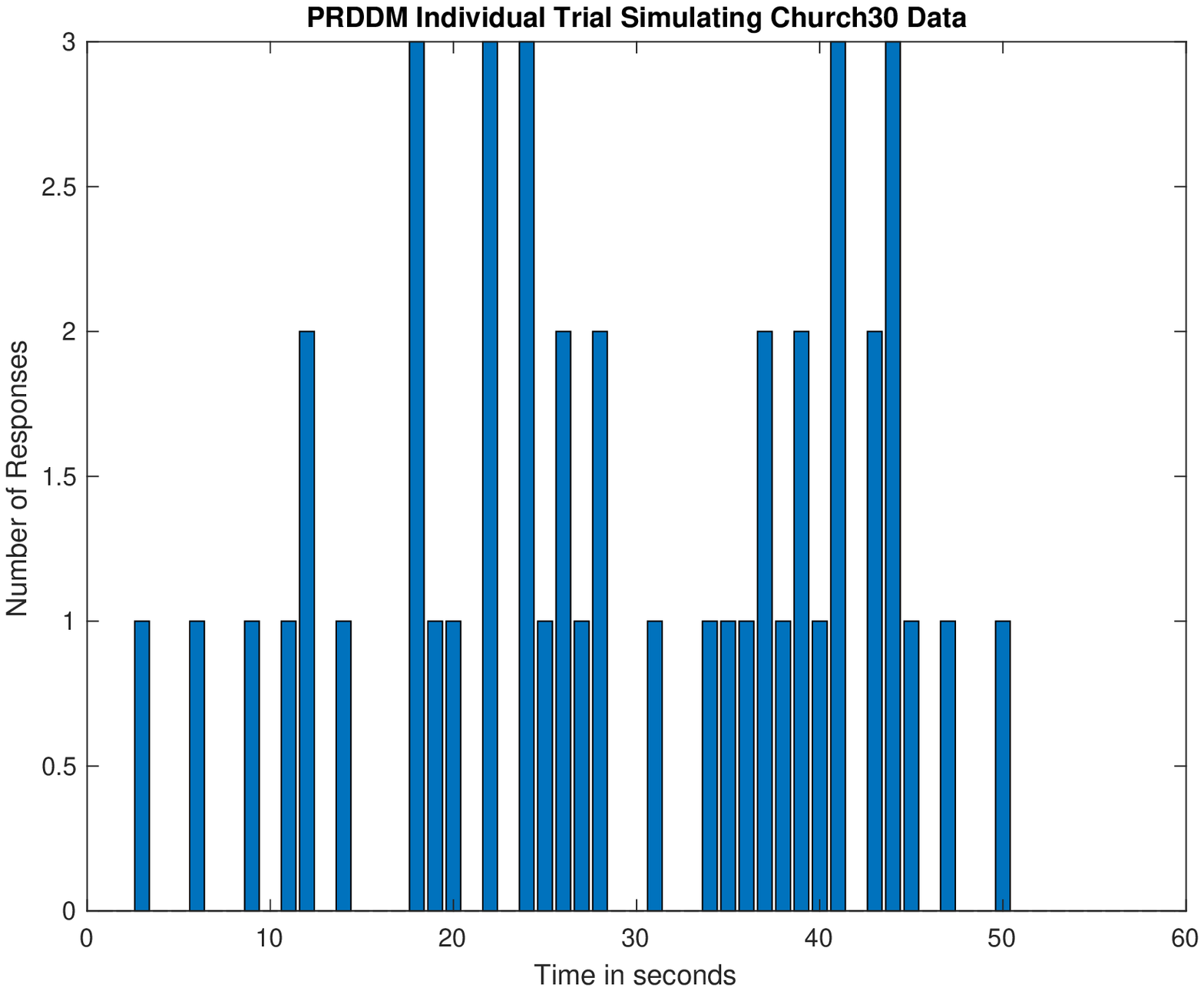}}

&\subfloat[TDDM] {\label{Fig_Church30_Ind_Trial_TDDM}
\includegraphics[width=0.29\textwidth]{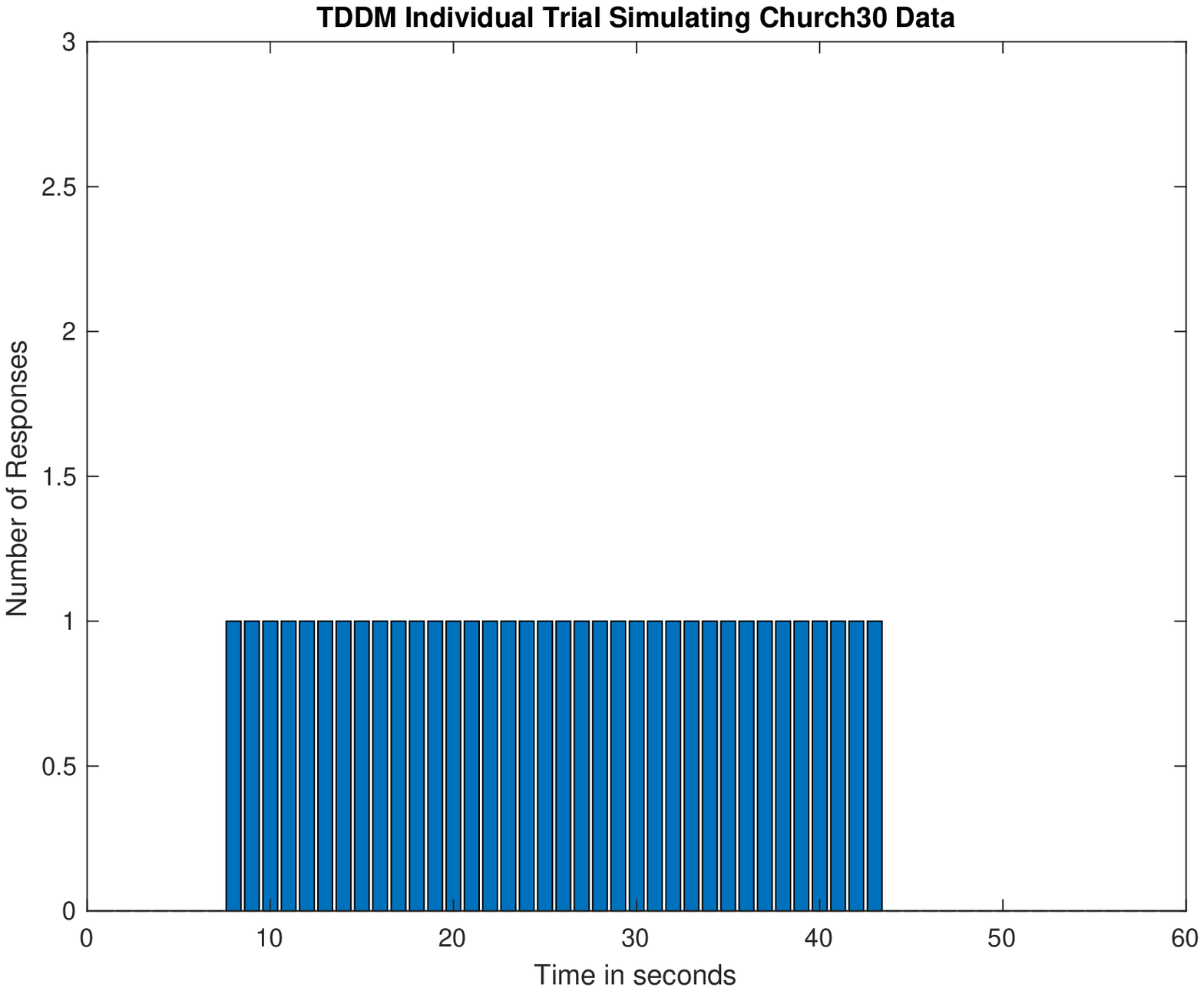}}

\\
\hline
\end{tabular}

\caption[PRDDM, and TDDM simulated individual trials for the Church30 data set]{\label{Fig_Church30_Ind_Trial_Compare} Response histograms of a single rat trial from (a) Church 1998 dataset (with FI = 30s) and a similar trial from the corresponding (b) PRDDM simulations and (c) TDDDM simulations respectively. While TDDM (c), like SET, is a pure break-run-break model, PRDDM (b) generates a more natural response (a) histogram.}
\end{figure}

\begin{figure} [t]
\centering

\begin{tabular}{|c|c|c|}
\hline

\subfloat[Balci] {\includegraphics[width=0.29\textwidth]{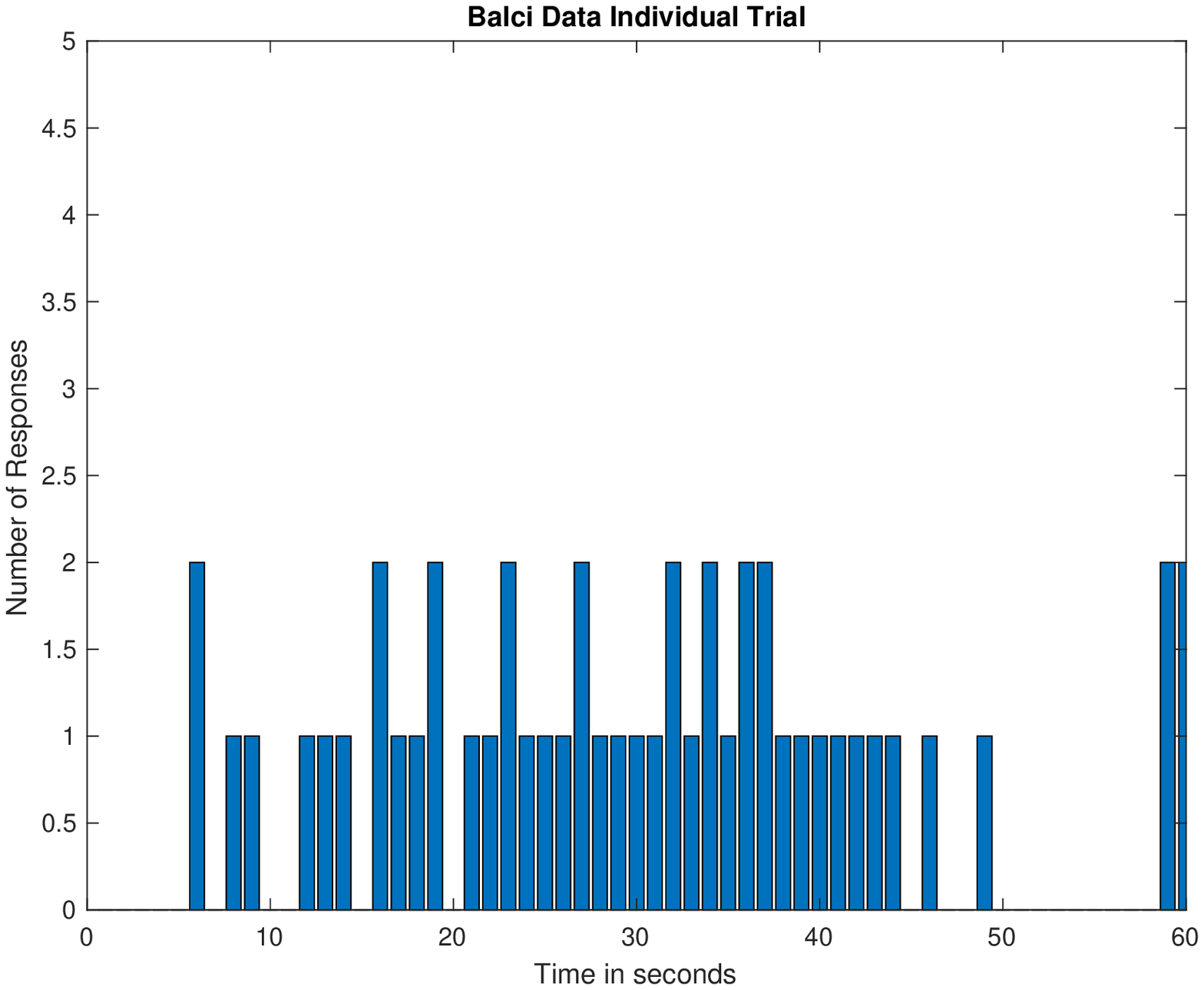}}
&
\subfloat[PRDDM] {\includegraphics[width=0.29\textwidth]{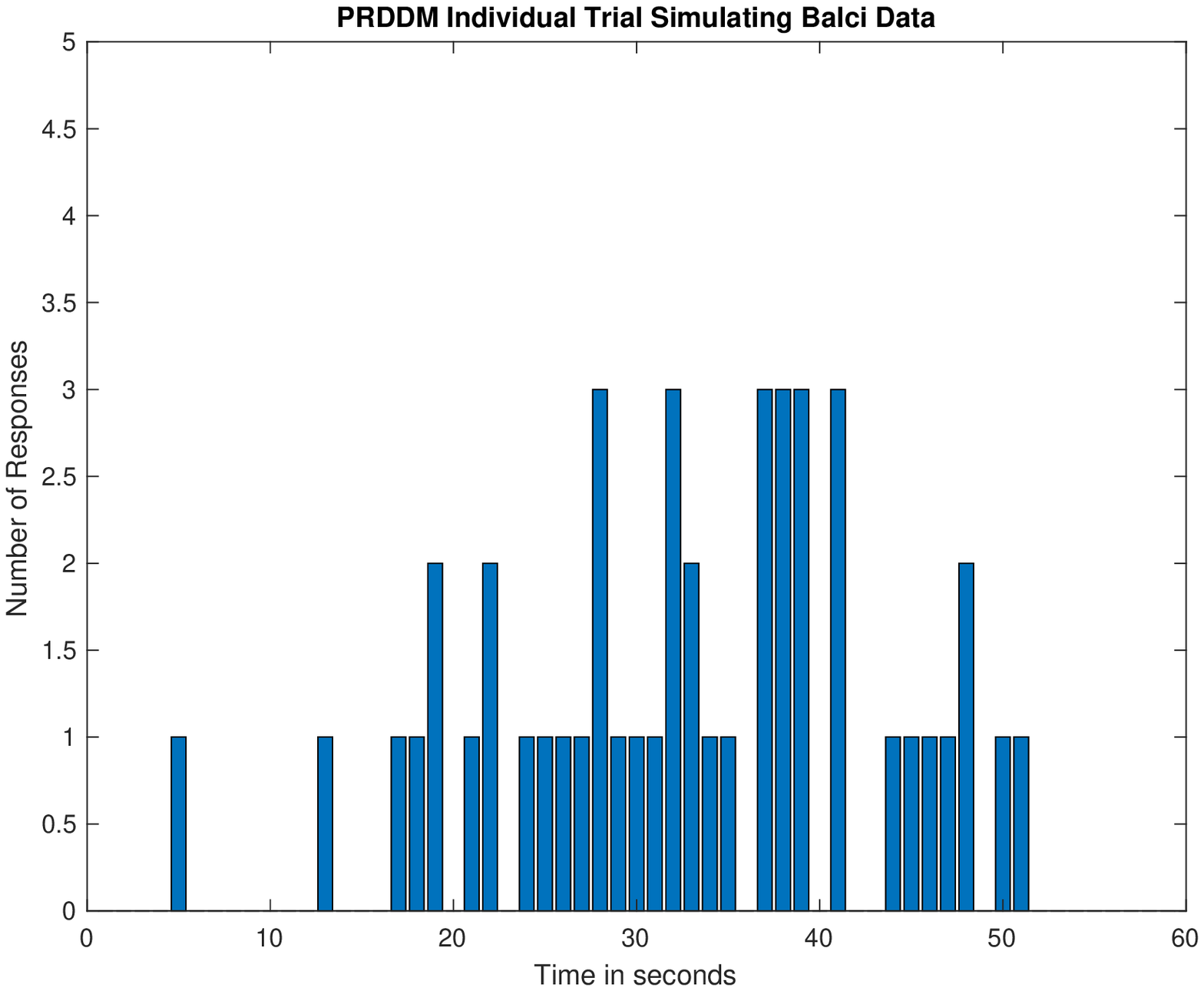}}
&
\subfloat[TDDM] {\includegraphics[width=0.29\textwidth]{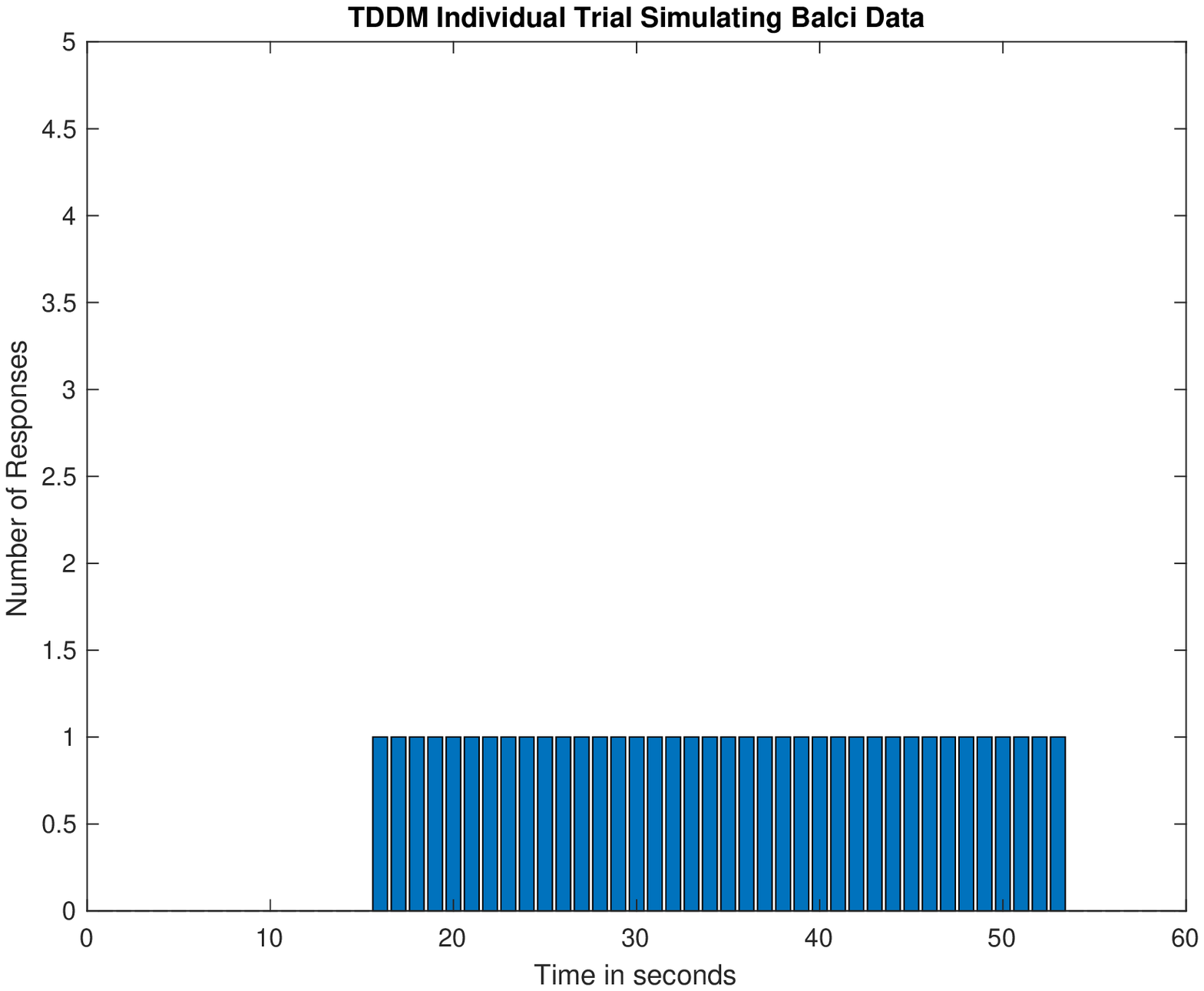}}

\\
\hline
\end{tabular}

\caption[PRDDM, and TDDM simulated individual trials for the Balci2009 data set]{\label{Fig_Balci_Ind_Trial_Compare} Response histograms of a single mouse trial from (a) Balci 2009 dataset (with FI = 30s) and a similar trial from the corresponding (b) PRDDM simulations and (c) TDDDM simulations respectively. While TDDM (c), like SET, is a pure break-run-break model, PRDDM (b) generates a more natural response (a) histogram.}
\end{figure}

At the individual trial level, TDDM and PRDDM, behave very differently. As shown in Fig.~\ref{Fig_Model_Process}, TDDM is a strict break-run-break model providing a continual response throughout the run and no response during the break. In contrast, PRDDM was developed to generate more realistic individual trials, comparable to experimental datasets, by using a probabilistic response throughout.  The differences between TDDM and PRDDM are easily observed in Fig.~\ref{Fig_Church30_Ind_Trial_Compare} and Fig.~\ref{Fig_Balci_Ind_Trial_Compare} in panel (b) (PRDDM) and panel (c)(TDDM).  The data contained in these two graphs is representative of the differences observed in all our simulations and their corresponding original experimental data.

As seen in Fig.~\ref{Fig_Church30_Ind_Trial_Compare} and Fig.~\ref{Fig_Balci_Ind_Trial_Compare} panel (a), the sample datasets do not correspond to a strict break and run pattern, but instead, they have an increasing response rate the closer time is to the subject's estimated time of reward. PRDDM is a closer match to this aspect of the individual trial than TDDM. PRDDM stochastic response also generates noisier and much more realistic individual trials than TDDM.

To further compare the models, we computed the mean and variance of the number of responses, response rate, and the peak time of each experiment and compared them to the corresponding simulations. The complete results can be found in Tables~\ref{Table_Balci2009_Individual} and \ref{Table_Church2000_Individual} in the Appendix. Since each model was used to reproduce each of the 14 different original experiments, we computed the mean and standard deviation of the absolute difference between the animal data and the simulations for each of those 6 statistics for both model. As it can be seen in Table~\ref{Table_MeanError_Individual}, PRDDM usually matches the statistics of the animal data better than TDDM (except for Mean Peak Time), but the standard deviation is too large to show significance. In short, sometimes PRDDM is better, sometimes TDDM is better, but on average, PRDDM seems better and is certainly no worse than TDDM. PRDDM also produces much more realistic individual trials.

In summary, qualitatively PRDDM provides a close match to the individual trial of all datasets when examined visually. In contrast, TDDM provides the expected break-run-break pattern, which is not similar to animal individual trials.  When the quantitative means and variances are considered, all models showed similar weaknesses; there was a significant underestimation of the variance of the experimental datasets.

Notably, the number of responses was fairly constant within each of the Church 1998 datasets no matter the FI. Specifically, with FI's of 30, 45, and 60, the number of responses were all approximately 41 per trial.

\begin{table}[!ht]
\centering
\begin{tabular}{|l|l|S[table-format=3.2]@{\;\( \pm \)} S[table-format=3.2]|}
\hline
Measure & Simulation & \multicolumn{2}{c|}{Abs Error} \\ \hline  
Mean Number of & PRDDM & 3.37 & 8.81 \\
Responses & TDDM & 12.52 & 14.67 \\ \hline
Mean Response Rate & PRDDM & 0.03 & 0.06 \\
 & TDDM & 0.08 & 0.07 \\ \hline
Mean Peak Time & PRDDM & 1.21 & 1.56 \\
 & TDDM & 0.86 & 1.18 \\ \hline
Variance on the & PRDDM & 151.61 & 174.14 \\
Number of Responses & TDDM & 192.25 & 219.46 \\ \hline
Varariance on the & PRDDM & 0.02 &  0.04 \\
Response Rate & TDDM & 0.02 & 0.04 \\ \hline
Variance on the & PRDDM & 49.22 & 103.13 \\
Peak Time & TDDM & 92.56 &  190.10 \\ \hline
\end{tabular}
\caption[Tabulated Mean Absolute Error for Individual Trial Statistics] {Mean and standard deviation of the absolute difference between the models' responses statistics and the animal statistics over all simulated experiments.}
\label{Table_MeanError_Individual}
\end{table}

\subsection{Start and Stop Analysis}

A hallmark of SET and TDDM models was their ability to reproduce trials' start and stop statistics as well as a number of their properties   \citep{church1994application, luzardo2017drift}:

\begin{itemize}
\item Positive correlation between start and stop,
\item Negative correlation between start and duration,
\item Positive correlation between duration and middle,
\item Coefficient of variation for the start larger than the stop.    
\end{itemize}

The complete correlations and CVs results for each simulation and experiment can be found in Tables~\ref{Table_Balci2009_StartStop_Corr}  \ref{Table_Church2000_StartStop_Corr} in the Appendix. As it can be seen, most values of TDDM and PRDDM are relatively close to the animal results.

\begin{table}[!ht]
\centering
\begin{tabular}{|l|l|S[table-format=3.2]@{\;\( \pm \)} S[table-format=3.2]|}
\hline
Measure & Simulation & \multicolumn{2}{c|}{Abs Error} \\ \hline
Mean Start Time ($S_1$) & PRDDM & 8.3 & 7.53 \\
  & TDDM & 1.57 & 1.84 \\ \hline
Mean Stop Time ($S_2$) & PRDDM & 4.06 & 3.53 \\
  & TDDM & 5.8 & 4.37 \\ \hline
Mean Mid-point ($M$) & PRDDM & 2.4 & 2.47 \\
 & TDDM & 2.41 & 1.65 \\ \hline
Mean Duration ($D$) & PRDDM & 11.89 & 10.99 \\
 & TDDM & 7.01 & 5.83 \\ \hline
Variance on $S_1$ & PRDDM & 31.85 & 33.94 \\
 & TDDM & 61.04 & 87.25 \\ \hline
Variance on $S_2$ & PRDDM & 39.08 & 38.85 \\
 & TDDM & 55.03 & 65.28 \\ \hline
Variance on $M$ & PRDDM & 36.45 & 46.84 \\
 & TDDM & 30.43 & 33.22 \\ \hline
Variance on $D$ & PRDDM & 67.73 & 132.23 \\
 & TDDM & 116.4 & 155.86 \\ \hline
\end{tabular}
\caption[Tabulated Mean and Standard Deviation of the Start and Stop Statistics] {Mean and standard deviation  of the absolute difference between the models start and stop statistics and the animal statistics over all simulated experiments.}
\label{Table_MeanError_StartStopMeanVar}
\end{table}

\begin{table}[!ht]
\centering
\begin{tabular}{|l|l|S[table-format=2.3]@{\;\( \pm \)} S[table-format=2.3]|}
\hline
Measure & Simulation & \multicolumn{2}{c|}{Abs Error} \\ \hline
$Corr(S_1,S_2)$ & PRDDM & 0.157 & 0.099 \\
 & TDDM & 0.110 & 0.081 \\ \hline
$Corr(S_1,M)$ & PRDDM & 0.099 &  0.065 \\
 & TDDM & 0.080 & 0.084 \\ \hline
$Corr(M,D)$  & PRDDM & 0.123 & 0.071 \\
  & TDDM & 0.137 & 0.107 \\ \hline
$CV_{Start}$ & PRDDM & 0.108 & 0.072 \\
 & TDDM & 0.063 & 0.025 \\ \hline
$CV_{Stop}$ & PRDDM & 0.018 & 0.019 \\
 & TDDM & 0.024 & 0.025 \\ \hline
\end{tabular}
\caption[Tabulated Mean and Standard Deviation of the Start and Stop Correlations] {Mean and standard deviation of the absolute difference between the models start and stop correlations and the animal correlations over all simulated experiments.}
\label{Table_MeanError_StartStopCorr}
\end{table}

Out of the 16 experiments, TDDM and PRDDM only failed one of the above four properties once, i.e., to have a positive correlation between the mid-point ($M$) and duration ($D$) on Balci 2009 dataset. The animal data also failed that property once in Church 2000 PI60-180 Noise and PI120 Noise datasets. All other correlation and CV from these $3 \times 16$ experiments follow these four properties (TDDM, PRDDM, and animal data included). 

Unlike SET and TDDM, PRDDM can reproduce those properties without explicit start and stop thresholds within the PA model. This supports the hypothesis that no internal break-run-break states are needed for a PA model to reproduce these statistics.

To further analyse which model is best, we compared the statistics of $S_1$, $S_2$, $M$ and $D$, $Corr(S_1,S_2)$, the correlations $Corr(S_1,M)$, $Corr(M,D)$, and the $CV_{Start}$ and  $CV_{Stop}$ of each models to the one of the corresponding animal dataset, as we did in the previous section. The results are summarized in Tables~\ref{Table_MeanError_StartStopMeanVar} and  \ref{Table_MeanError_StartStopCorr}. While TDDM tends to be closer in mean $S_1$, and $D$, PRDDM tends to more closely match the natural variance of $S_1$, $S_2$, and $D$. TDDM is a slightly better match at two of the three correlations, each model is better at one of the CVs. Overall, if we were to perform a paired $t$-test with Bonferroni correction, only the mean $S_1$ (start time) would be significantly different, with TDDM being a closer match to the animal data. Otherwise, both models are sensibly as good as one another.

\begin{figure}[!htbp]
\centering

\begin{tabular}{|c|c|}
\hline
\subfloat[Start Church 1998 PI 30] {\includegraphics[height=0.175\textheight]{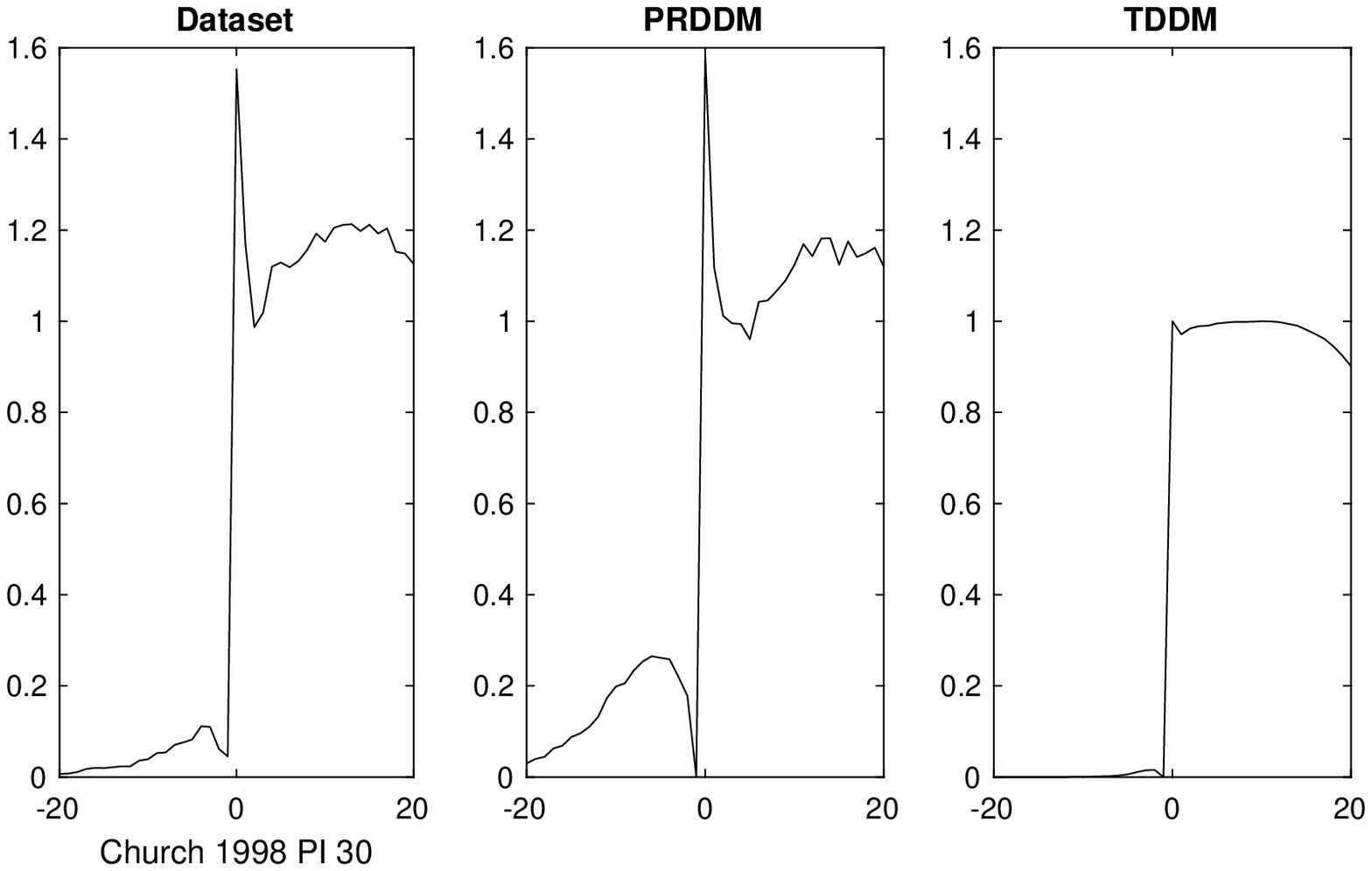}}
&
\subfloat[Stop Church 1998 PI 30] {\includegraphics[height=0.175\textheight]{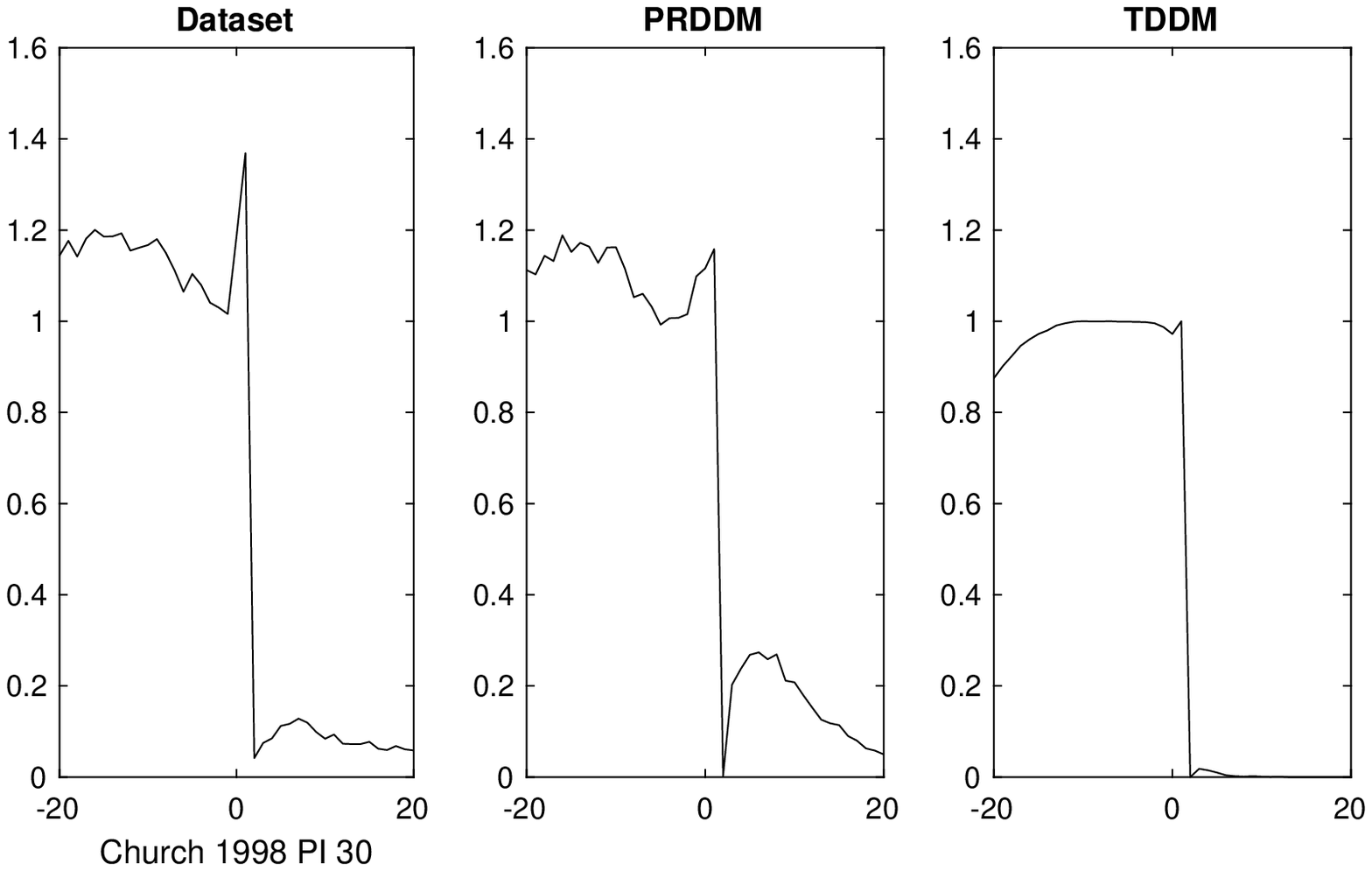}}
\\
\hline
\subfloat[Start Church 2000 30 Light] {\includegraphics[height=0.175\textheight]{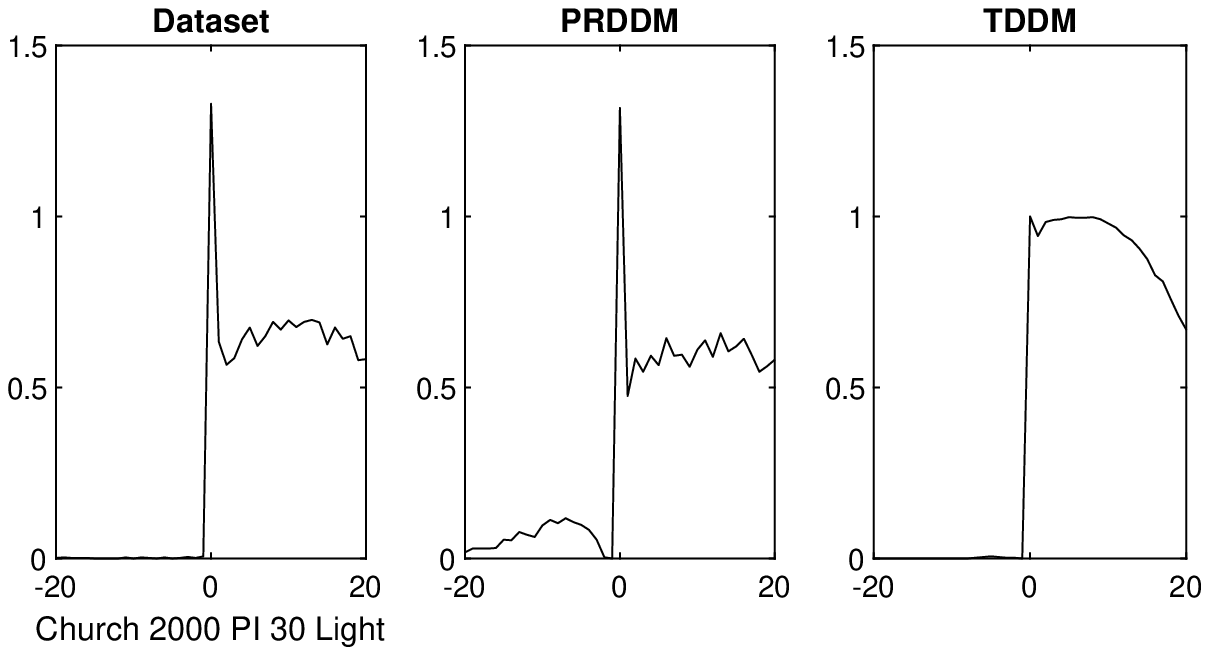}}
&
\subfloat[Stop Church 2000 30 Light] {\includegraphics[height=0.175\textheight]{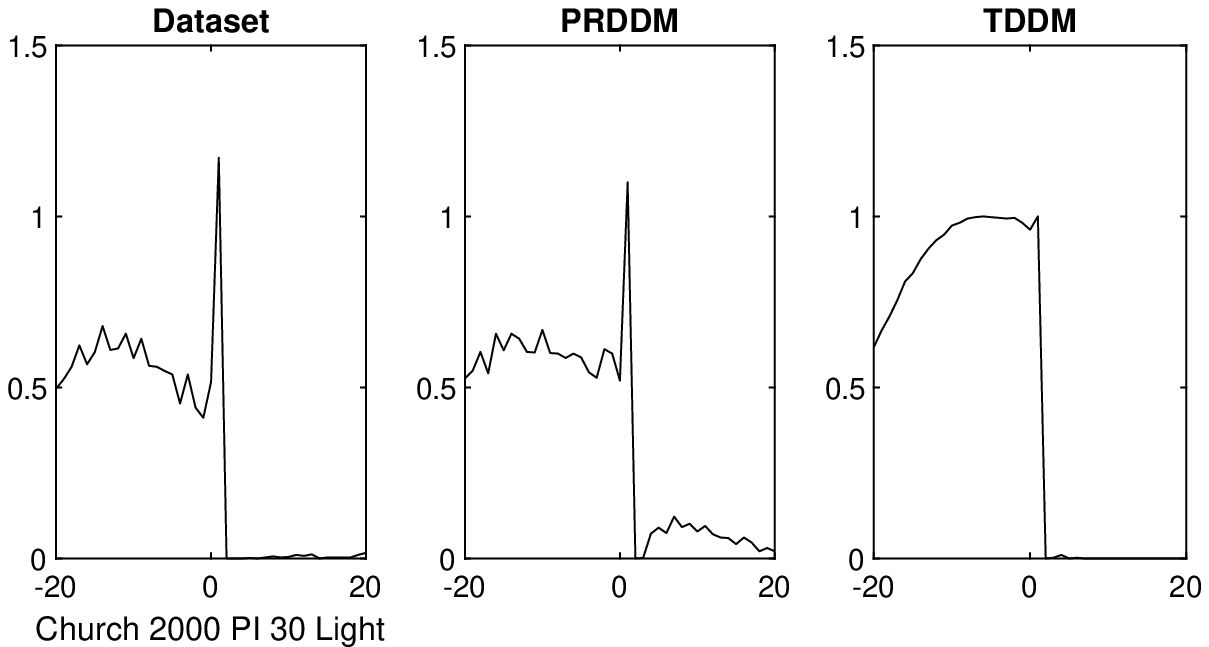}}
\\
\hline
\subfloat[Start Church 2000 60 180 Light] {\includegraphics[height=0.175\textheight]{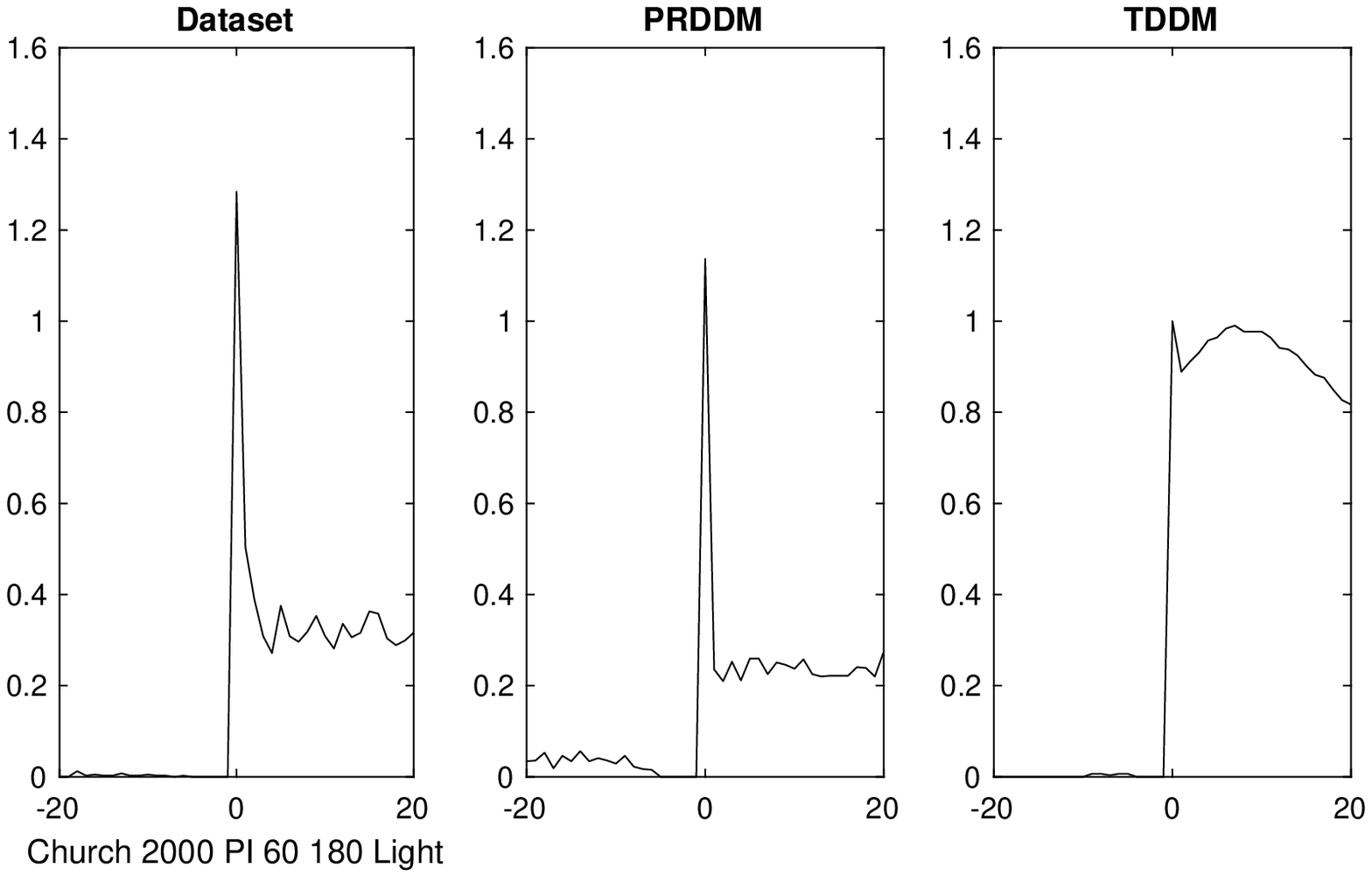}}
&
\subfloat[Stop Church 2000 60 180 Light] {\includegraphics[height=0.175\textheight]{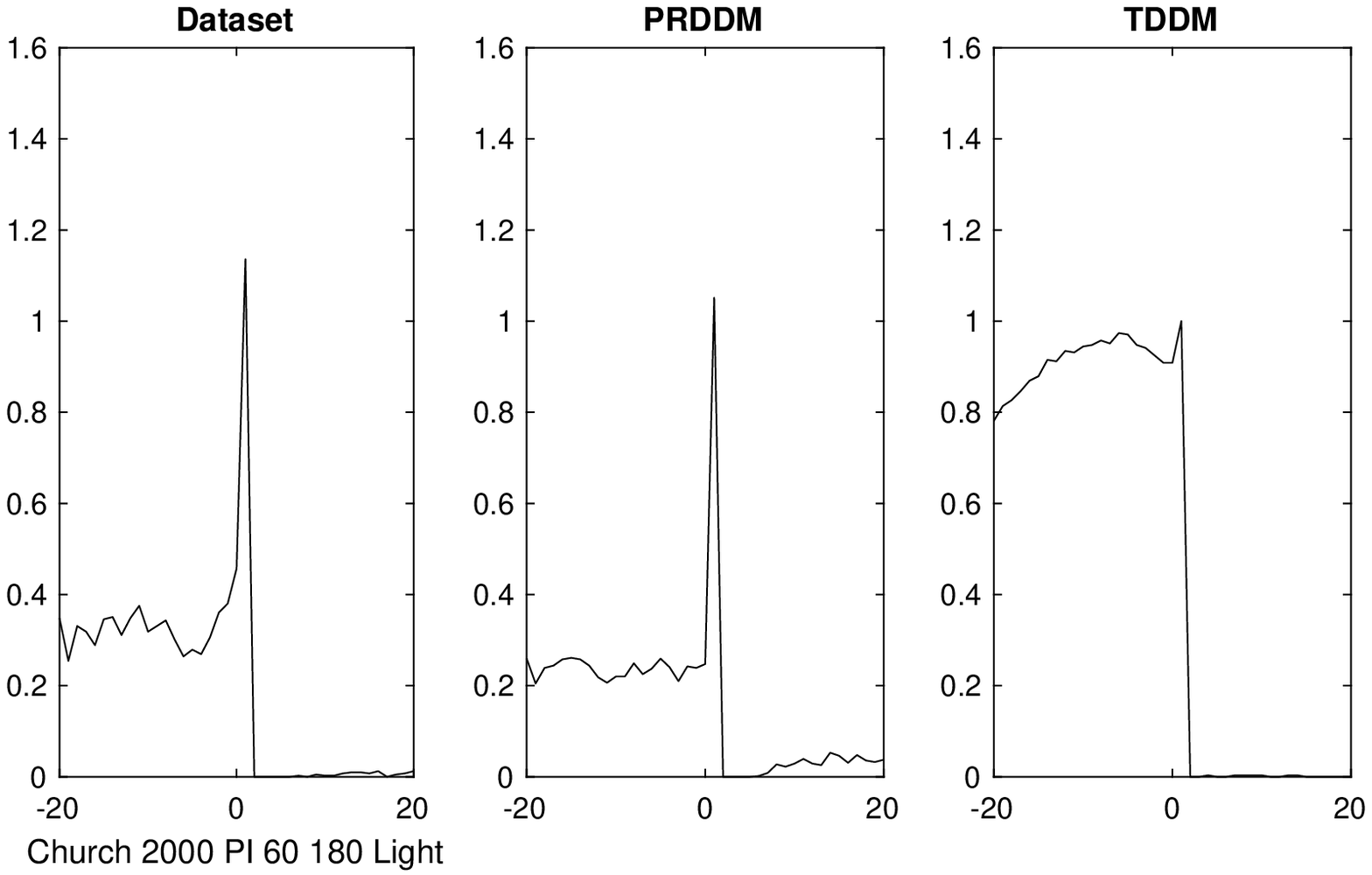}}
\\
\hline
\end{tabular}
 
\caption[Church 2000 PI30, Church 2000 PI 60 and Church 1998 PI 30 Aligned on Start Stop]{\label{Fig_Balci_StartStop} This figure graphs the Church 2000 PI 30, Church PI 30 Light, and Church 1998 PI 60 Light datasets aligned on the start time (a and c) and the stop time (b and d).  From left to right, you see the dataset, the PRDDM simulation and finally the TDDM simulation.}
\end{figure} 

Finally, in \citet{cheng1993analysis} and \citet{church1994application}, the authors align individual trials on the start point and separately on the stop point. This showed an abrupt change in response rate at that point.  We have replicated that analysis with all of our datasets and related simulations. Despite PRDDM having no internal start or stop decision or inflection points, PRDDM showed an abrupt change in response rate similar to the graphs in \citet{cheng1993analysis} and \citet{church1994application} on each dataset we analyzed. Graphs representative of our results can be seen in Fig.~\ref{Fig_Balci_StartStop}. The fact that the step is visible on both, the animal datasets, and PRDDM, suggests that this may very well be an artifact of the analysis procedure. Since the procedure that searches for a start and a stop assumes their existence, it may find the most step-like decision point. When averaged over many trials, this effect could be amplified, as demonstrated by PRDDM results. There are still small visible differences between the dataset and PRDDM before and after the jump in response rate which may be caused by the strong linear constraint within PRDDM. Nevertheless, PRDDM provides a very good match to the animal data, despite its lack of start and stop thresholds.

In short, TDDM and PRDDM are equally strong across all start and stop statistics and provide the expected sign and rough magnitude for the correlations.  In general, both models did not consistently capture the variance of the experimental datasets.  This is likely due to not modeling individual subjects but rather modeling the trial's population as a whole with a single distribution, or caused by other sources of variance in animals in general. Despite not having an explicit start and stop decision or inflection point, PRDDM was able to reproduce the classic step-like response rate seen in animal datasets when trials are aligned on the start or stop.

\subsection{Average Behaviour}

We first assessed the average response curves qualitatively through assessing their graphs.  We followed this with a qualitative assessment comparing the experimental response distribution to the simulated response distributions using both $AIC_c$ and ICC.

\begin{figure}[!htbp]
\centering

\begin{tabular}{|c|c|}
\hline
\subfloat[PRDDM] {\includegraphics[width=0.45\textwidth]{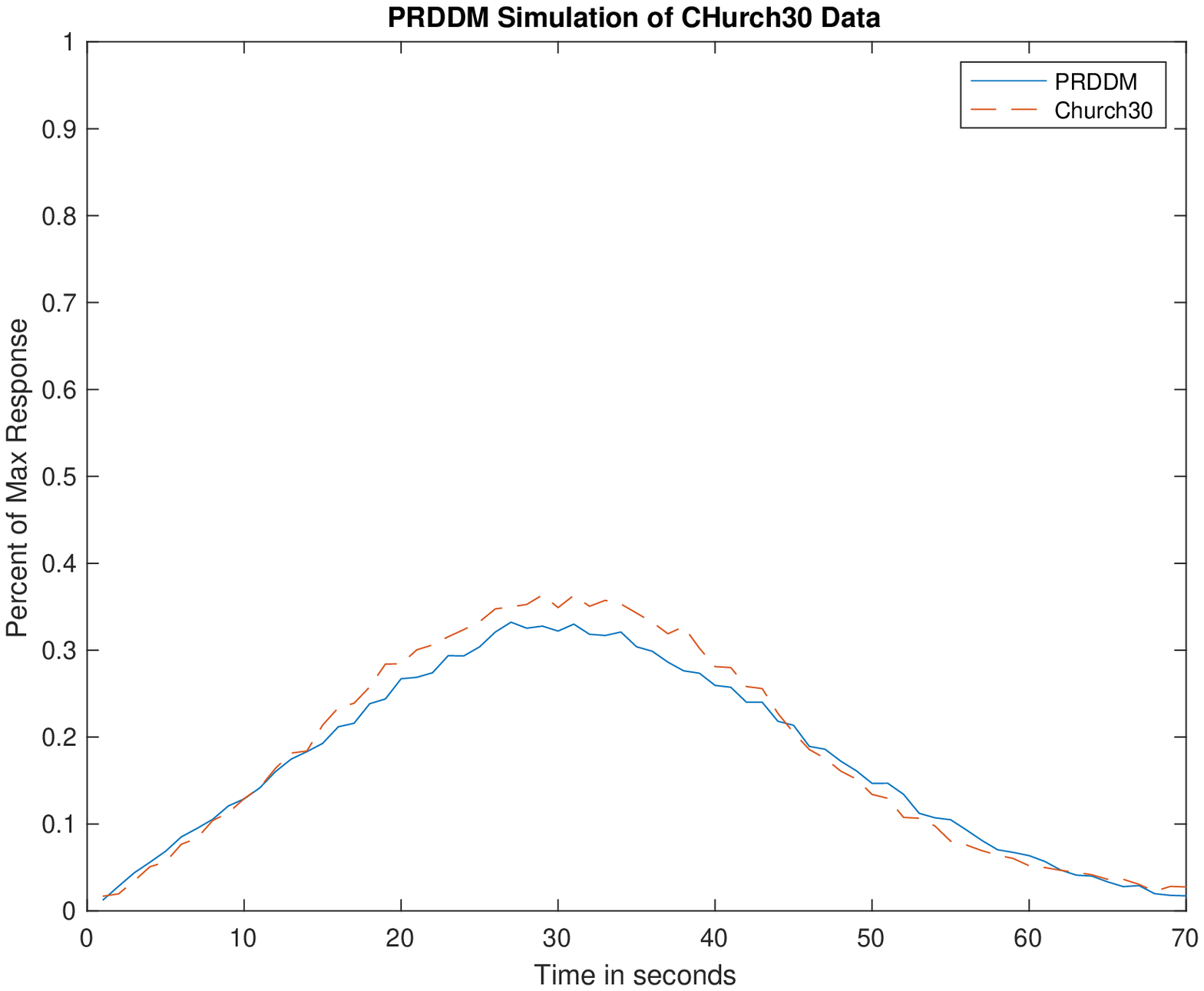}
\label{Fig_Church30_Resp_PRDDM}}
&
\subfloat[TDDM] {\includegraphics[width=0.45\textwidth]{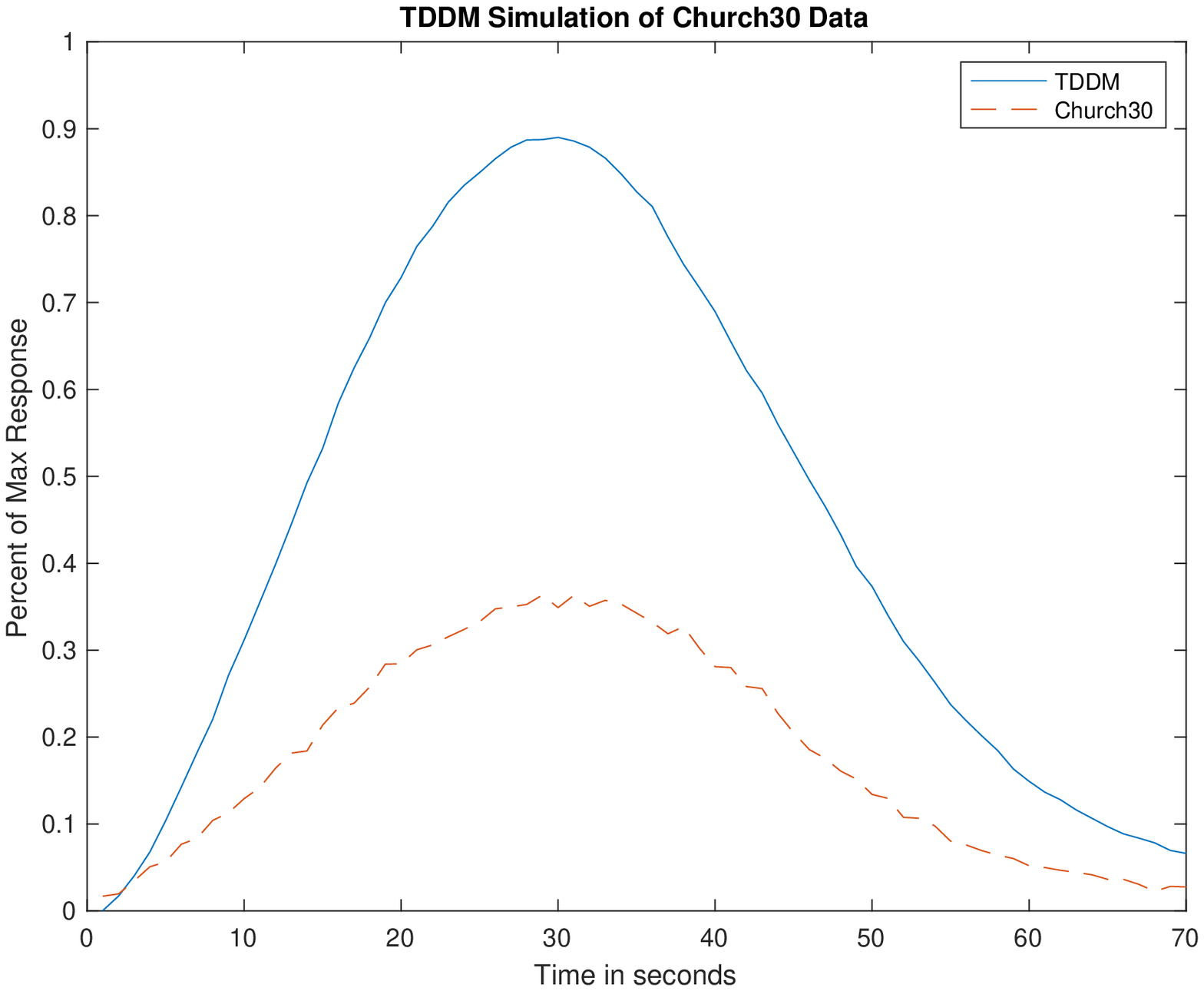}}

\\
\hline
\subfloat[PRDDM] {\includegraphics[width=0.45\textwidth]{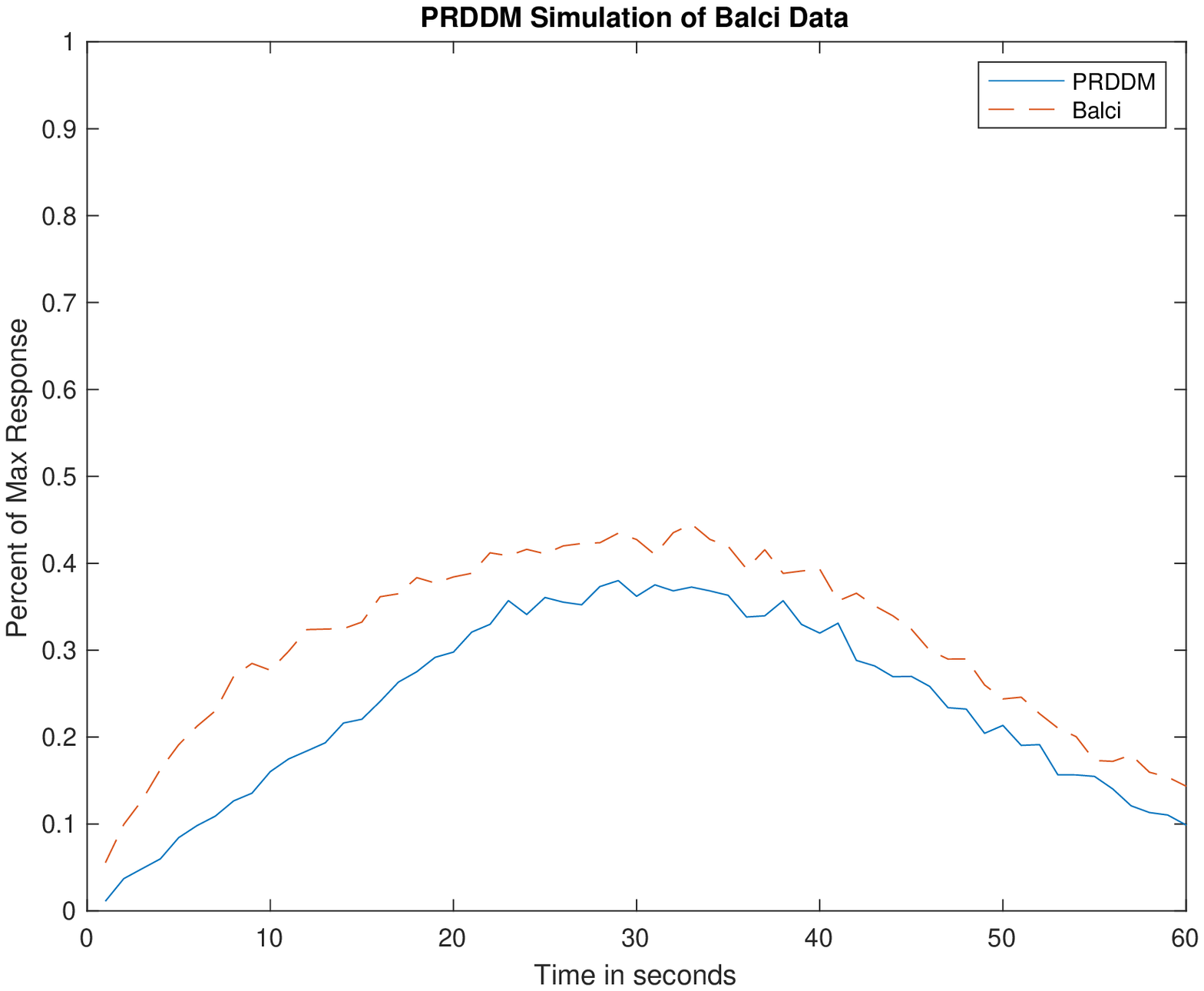}}
&
\subfloat[TDDM] {\includegraphics[width=0.45\textwidth]{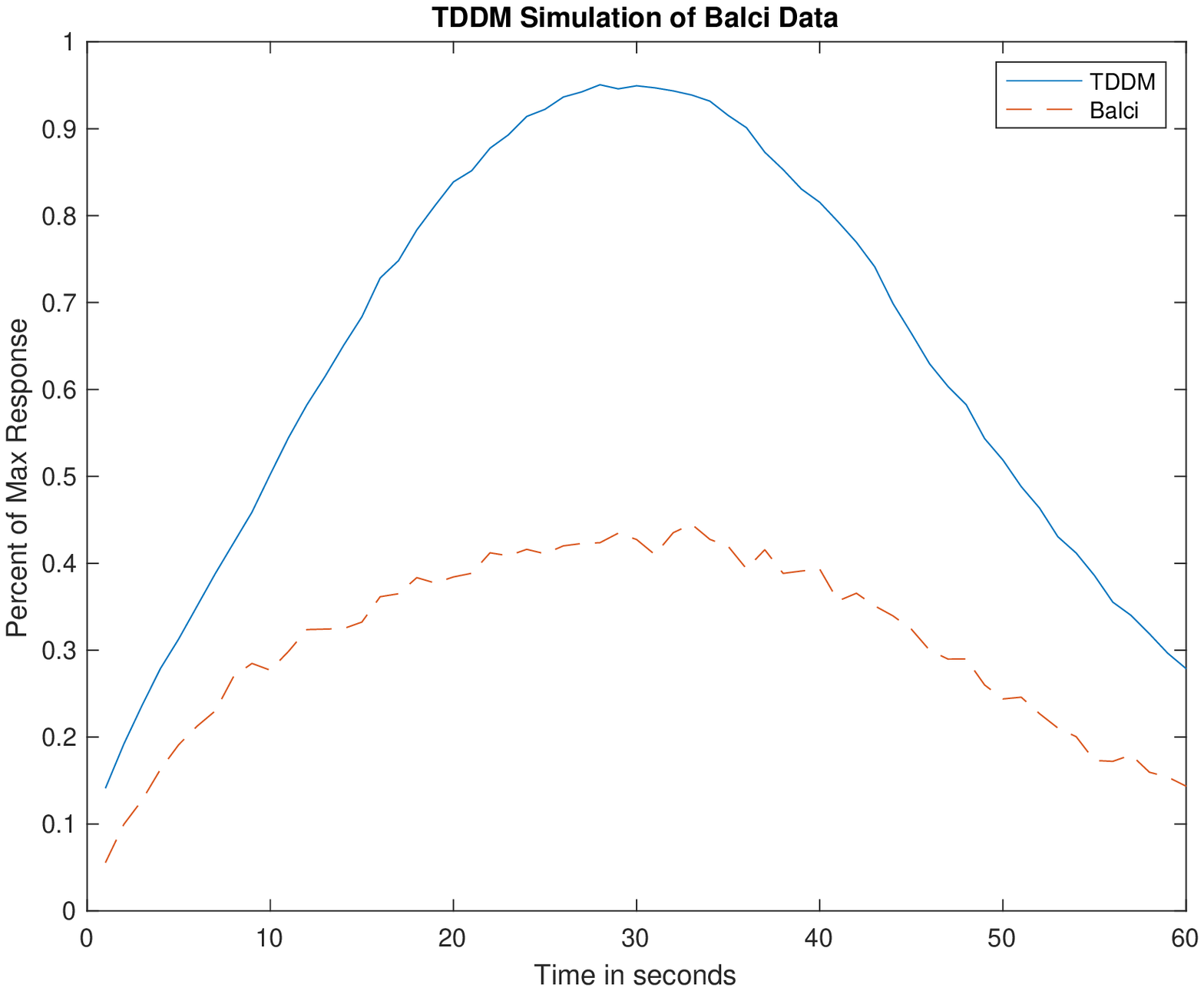}}

\\
\hline
\end{tabular}
 
\caption[PRDDM and TDDM average response curves for the Church30 data set]{\label{Fig_Response_Curves_Church30} These graphs display the response curves of PRDDM and TDDM simulations of the Church30 (graphs a and b) and the Balci2009 (graphs c and d) dataset compared to the experimental data.  Visually, it is clear that PRDDM is a better match for the Church30 and the Balci2009 dataset.  The difference is due to the constant response of TDDM during the run period.  The constant response, when averaged, is a response rate of 100 percent at the peak as the subject is always responding during the peak.  This differs significantly from PRDDM as its response is probabilistic during the peak.  The $y$-axis is percent response, and the $x$-axis is bins in second.}
\end{figure}

As shown in Fig.~\ref{Fig_Response_Curves_Church30} there is a marked difference between the PRDDM simulation and the TDDM simulation.  PRDDM simulates both the shape and the magnitude of the response distribution while TDDM only simulates the shape.  This difference between TDDM and PRDDM is caused by the strict break-run-break nature of TDDM.  As TDDM has a constant response rate during the run period, and those periods all align with the peak time, TDDM will have a percentage of max response rate of 100\% during the peak.  Whereas mice have a variable response rate during the peak period and as seen have a percentage of max response rate of approximately 30\%. PRDDM reproduces this property. 

The quantitative assessment was based on $AIC_C$ and ICC values for the simulated distributions when compared to the experimental distributions.  For all datasets, $AIC_C$ showed that PRDDM was a significantly better fit than TDDM.  While ICC also showed that PRDDM was a significantly better fit, it also showed that PRDDM produced good (ICC Value > 0.75) or excellent (ICC Value > 0.9) fit for all datasets except for the Church 2000 PI 120 datasets and that the fit for the PI 120 datasets would be considered moderate (ICC Value > 0.5).

\begin{table}[!ht]
\centering
\begin{tabular}{|l|l|l|l|l|}
\hline
Paper & Dataset & Simulation & AICC & ICC \\ \hline
Balci 2009 & PI30 & PRDDM & -4791.68 & 0.802 \\
Balci 2009 & PI30 & TDDM & -2674.51 & 0.289 \\ \hline
Church 1998 & PI30 & PRDDM & -25936.11 & 0.989 \\
Church 1998 & PI30 & TDDM & -10971.23 & 0.271 \\ \hline
Church 1998 & PI45 & PRDDM & -22885.39 & 0.966 \\
Church 1998 & PI45 & TDDM & -9422.52 & 0.190 \\ \hline
Church 1998 & PI60 & PRDDM & -17891.24 & 0.933 \\
Church 1998 & PI60 & TDDM & -6573.32 & 0.286 \\ \hline
Church 2000 & PI15 Noise & PRDDM & -4906.40 & 0.945 \\
Church 2000 & PI15 Noise & TDDM & -1688.57 & 0.451 \\ \hline
Church 2000 & PI15 Light & PRDDM & -4470.06 & 0.918 \\
Church 2000 & PI15 Light & TDDM & -1331.99 & 0.584 \\ \hline
Church 2000 & PI30 Noise & PRDDM & -4946.07 & 0.876 \\
Church 2000 & PI30 Noise & TDDM & -1949.62 & 0.460 \\ \hline
Church 2000 & PI30 Light & PRDDM & -5748.30 & 0.957 \\
Church 2000 & PI30 Light & TDDM & -2002.49 & 0.303 \\ \hline
Church 2000 & PI60 90 Noise & PRDDM & -4673.51 & 0.760 \\
Church 2000 & PI60 90 Noise & TDDM & -849.44 & 0.217 \\ \hline
Church 2000 & PI60 90 Noise & PRDDM & -4949.28 & 0.835 \\
Church 2000 & PI60 90 Noise & TDDM & -885.15 & 0.255 \\ \hline
Church 2000 & PI60 180 Noise & PRDDM & -5189.69 & 0.834 \\
Church 2000 & PI60 180 Noise & TDDM & -1668.25 & 0.133 \\ \hline
Church 2000 & PI60 180 Light & PRDDM & -4870.80 & 0.822 \\
Church 2000 & PI60 180 Light & TDDM & -955.59 & 0.171 \\ \hline
Church 2000 & PI120 Noise & PRDDM & -4540.23 & 0.659 \\
Church 2000 & PI120 Noise & TDDM & -932.69 & -0.039 \\ \hline
Church 2000 & PI120 Light & PRDDM & -4087.52 & 0.518 \\
Church 2000 & PI120 Light & TDDM & -847.30 & -0.118 \\ \hline
\end{tabular}
\caption[Tabulated AICC and ICC Values] {AICC and ICC values for all datasets.  More negative AICC values indicate closer datasets.  For ICC values closer to 1 indicate a closer match between datasets.}
\label{Table_Response_AICC_ICC}
\end{table}

As seen in Fig.~\ref{Fig_Response_Curves_Church30}, PRDDM qualitatively provides a better fit to the experimental dataset.  This was further confirmed with the quantitative assessment. The values in Table~\ref{Table_Response_AICC_ICC} show a marked difference between PRDDM and TDDM.  In particular, the fit of PRDDM to the data was significantly better compared to TDDM.

\subsection{Summary}

PRDDM and TDDM, while both being PA models, simulate the PI procedure with two different methods. TDDM is a two-state model with thresholds, implementing SET, explicitly defining break-run-break states. In contrast, PRDDM is a  probabilistic model whose response varies on the estimated elapsed time, without an explicit start or stop thresholds. Instead, they share an underlying accumulator that gathers evidence as time passes and the stimulus is present. The critical difference is that TDDM changes state to a high response when the accumulator exceeds a threshold, while PRDDM responds stochastically using the same underlying accumulator.  

Through an in-depth analysis of data from 14 different animal experiments, we used both PRDMM, and TDDM, to generate matching data by estimating their parameters from each animal experiment. The key results are:
\begin{itemize}
\item PRDDM produce much more natural-looking individual trials (response histograms) than TDDM in Fig.~\ref{Fig_Balci_Ind_Trial_Compare},
\item Both models produce quantitatively similar start-stop statistics properties in agreement with the animal data. 
\item For the average of the trial (average response curve), see Fig.~\ref{Fig_Response_Curves_Church30}, PRDDM is a better fit the animal data than TDDM.
\end{itemize}

Moreover, the trial start and stop statistics properties are correctly reproduced by PRDDM, despite the absence of explicit start and stop thresholds. While previous authors claimed it was impossible for a PA model to have only one threshold to reproduce these properties, they only looked at a combination of accumulator and thresholds under the assumption that there was an explicit start and stop defining the break-run-break pattern. They did not consider that the animal may actually not break down their response that way and that the accumulator could have been used directly to produce a probabilistic response. The results of PRDDM simulations suggest that it is possible that the animal process may not include a break-run-break pattern at all. Furthermore, it shows that this can be achieved with a PA model.

\section{Discussion}

\subsection{Pacemaker-Accumulator Models and Thresholds}

Pacemaker-accumulator models, such as SET \citep{gibbon1977scalar, church1994application}, have used thresholds to produce within the model explicit start and stop for the period of high rate responding in the PI procedure.  Through a detailed analysis of the start and stop times, it has been shown that two thresholds most closely match the experimental results \citep{gibbon1990representation,church1994application,luzardo2017drift}.  

PRDDM uses a noisy internal clock with a probabilistic response based on the estimated likelihood of event occurrence to simulate animal behaviour on the PI procedure to a high degree of accuracy.  This noise within the internal clock is sufficient to generate the observed start and stop times, provided a stochastic response mechanism is given.  In addition, it questions the break-run-break response pattern of the PI procedure and the corresponding analysis methods.  As PRDDM has no start or stop thresholds or boundaries, it supports that the response could continuously vary with the subject's estimated likelihood of event occurrence, as proposed in packet theory \citep{kirkpatrick2002packet,kirkpatrick2003tracking}.

We showed that the observed break and run pattern, start and stop descriptive statistics, individual trials, and the average behaviour can all be generated without start and stop thresholds and with a model where the response rate varies with the estimated likelihood of reward.  This directly contrasts with other pacemaker-accumulator models in that it is not a two-state system corresponding to break-run-break.  This suggests a different relationship between the internal clock and decision making, closer to packet theory \citep{kirkpatrick2002packet,kirkpatrick2003tracking}.  In particular, in PRDMM, the internal clock is used to estimate the likelihood of reward and based on that estimate, a decision to respond is made.

\subsection{PRDDM}

PRDDM successfully integrates key features from SET, TDDM \citep{simen2011model, rivest2011, simen2013timescale, luzardo2017drift} and Packet Theory \citep{kirkpatrick2002packet,kirkpatrick2003tracking}, resulting in a simple model of interval timing for a single stimulus.  From SET, the basic architecture remains the same, with a pacemaker-accumulator and reference memory.  The reference memory in SET corresponds to the drift rate in TDDM and PRDDM. Moreover both SET and PRDDM share the same reflective boundary. PRDDM was heavily influenced by TDDM and used a time-scaled DDM \citep{ratcliff1978theory} to implement the pacemaker-accumulator, which encodes the time interval in the drift rate and the estimate of the fraction of elapsed time in the accumulation process.  However, unlike SET and TDDM, PRDDM has no explicit start and stop thresholds. Instead, and similarly to Packet Theory, the response pattern during a single trial is determined by the subject's estimated likelihood of reward occurrence at each time step using the accumulator (Eq.~\ref{ProbResp}), thus eliminating the start and stop thresholds and the strict break-run-break pattern of SET and TDDM.  Overall, this produced an accurate and simple model of interval timing, both increasing the accuracy and decreasing the overall complexity.

\subsubsection{Linear assumption}

The central assumption of PRDDM is a linear relationship between the accumulator and the probability of response.  This assumption produces individual trials that tend to over concentrate responses at the estimated time of reward with a maximum number of responses in a bin being higher than the experimental maximum.  This is caused by the linear assumption.  

In the start and stop analysis, PRDDM consistently had minor differences in that starting late, stopping early, and a smaller spread than the experimental data; see Tables \ref{Table_Balci2009_StartStop_MeanVar}--\ref{Table_Church2000_StartStop_Corr} in the appendix.  This was likely caused by the assumed linear relationship between response and event likelihood.  In particular, the linear assumption ensures that the subject responds at maximum likelihood for a single time step and overly concentrates the response at the subject's estimated time of reward.  Recall the linear comparison to peak aligned trials in Sec.~\ref{Ch_PI_Model_Parameters}.  Therefore, we suggest that the minor differences in shape are due to a physical limitation of the subject's ability to respond.

Note that it is very possible that non-linearity in the response rate exists in animals and the current results do not contradict this possibility. It would be simple to add a non-linear relationship by changing Eq.~\ref{ProbResp} with 
\begin{equation}
\label{ProbResp2}
P(R=1 \mid t) = \tau f(x(t)),
\end{equation}
where $f(\cdot)$ represents the non-linearity. A similar approach has been used in \citet{luzardo2017rescorla} with the same TDDM PA model to replicate a wide range of conditioning behaviors successfully. While adding an extra non-linearity could further improve the current results, it has not been necessary to reproduce the animal data and other models.

\subsection{Break-Run-Break}

PRDDM produced the start and stop statistics \citep{church1994application, luzardo2017drift} of the break-run-break pattern to the same degree of accuracy or higher than current PA models.  These statistics are generally produced using thresholds for start and stop of the run state for pacemaker-accumulator models.  In the case of PRDDM, thresholds are not used to determine a start and stop of responding, rather the response rate varies directly with the accumulator value. While the break-run-break pattern seems observable in the binned data (when using a small enough bin size), it does not mean that the animal follows this internal process. As we showed (and as earlier suggested by Packet theory \citep{kirkpatrick2002packet, kirkpatrick2003tracking}), the underlying process may not have such break-run-break states at all. However, this does not devalue it, the break-run-break analysis, as we showed, reports similar results when using our non break-run-break (PRDDM) simulated trials or our break-run-break trials (TDDM).

Some other models do not necessarily assume start and stop \citep{machado1997learning,oprisan2011modeling,hasegawa2015model, raphan2019modeling}. However, they do not model individual trials either. Instead, they model average response curves. The only exception we found is \citet{varella2019model}. In this model, a network goes through a sequence of states. When the associate weight of the current state is below a threshold $0 < l < 1$, the model responds with a low rate, while when it is higher, it responds with a high rate. Unfortunately, the paper only compares the start and stop times distributions to rats, without an in-depth analysis of the mid-points, duration, correlations, or CVs.

\subsection{Extensions to other timing tasks}

It remains open to see how TDDM or PRDDM could explain other timing tasks. For example, in the bisection task, subjects are trained to two time intervals, on long and one short.  The subject is then exposed to varied short and long durations that have not been trained. If the subject correctly chooses short or long for the duration it is rewarded.  In a recent paper, \citet{balci2014decision} showed that TopDDM could be adapted to it using a sequential diffusion model. The same method should apply to PRDMM using a similar approach.

Other tasks, like differential reinforcement of low rates of responding (DRL) \citep{skinner1938behavior} seems to be more problematic. In DRL, the subject must wait to respond until after the minimum wait time.  If they respond prior to the wait time, the timer on reward is reset. One way this could play out, is that every early response restart the environment clock, making any reward extremely delayed. Depending on whether the model resets its accumulation or not, it could then learn a very long timing interval, leading to a very low response rate at all points in time during the standard interval it has to wait. But this would also lead to remarkably high variance on the first response timing. Further analysis would be required to see if this is possible. But that task could undoubtedly challenge the PRDDM model. So the solution as to whether thresholded or probabilistic response is better may not come from trying to overanalyze co-variance and other statistics on a specific task, but rather on developing a more complete model that encompasses different tasks with different properties or structures.
\\
\section{Conclusion}

In this paper, we proposed a new PA model of the PI procedure. Instead of assuming a start and stop decision threshold with an either on or off response rate, assumes a response rate directly proportional to elapsed time. While most analyses of PI data assume a break-run-break pattern, the model shows that it is not necessarily part of the animal process. A start/stop free model can reproduce animal data as well as SET and TDDM, including per trial start and stop analysis.

\section*{Acknowledgment}
We would like to thank Fuat Balci for making his data available.

\newpage

\begin{appendices}
\begin{landscape}

\section{Tables of Detailed Results}

\begin{table}[!ht]
\begin{tabular}{|lllllllll|}
\hline
Paper & Dataset & Simulation & \begin{tabular}[c]{@{}l@{}}Mean Num \\ Resp\end{tabular} & \begin{tabular}[c]{@{}l@{}}Mean Resp \\ Rate\end{tabular} & \begin{tabular}[c]{@{}l@{}}Mean Peak \\ Time\end{tabular} & \begin{tabular}[c]{@{}l@{}}Var Num \\ Resp\end{tabular} & \begin{tabular}[c]{@{}l@{}}Var Resp \\ Rate\end{tabular} & \begin{tabular}[c]{@{}l@{}}Var Peak \\ Time\end{tabular} \\ \hline
Balci 2009 & PI30 & PRDDM & 60.83 & 1.01 & 34.25 & 119.17 & 0.03 & 39.29 \\
Balci 2009 & PI30 & TDDM & 41.86 & 0.70 & 33.53 & 87.51 & 0.02 & 35.97 \\
Balci 2009 & PI30 & Data & 55.21 & 0.92 & 30.64 & 511.92 & 0.14 & 51.84 \\ \hline
Church 1998 & PI30 & PRDDM & 42.94 & 0.57 & 31.32 & 98.99 & 0.02 & 40.02 \\
Church 1998 & PI30 & TDDM & 33.37 & 0.44 & 30.97 & 117.47 & 0.02 & 47.60 \\
Church 1998 & PI30 & Data & 41.97 & 0.56 & 31.75 & 736.26 & 0.13 & 60.17 \\ \hline
Church 1998 & PI45 & PRDDM & 40.50 & 0.36 & 45.22 & 93.49 & 0.01 & 90.78 \\
Church 1998 & PI45 & TDDM & 41.38 & 0.37 & 46.13 & 196.94 & 0.02 & 91.72 \\
Church 1998 & PI45 & Data & 40.25 & 0.36 & 45.89 & 350.56 & 0.03 & 109.88 \\ \hline
Church 1998 & PI60 & PRDDM & 74.50 & 0.50 & 62.58 & 245.97 & 0.01 & 154.69 \\
Church 1998 & PI60 & TDDM & 48.07 & 0.32 & 61.14 & 348.80 & 0.02 & 136.55 \\
Church 1998 & PI60 & Data & 40.94 & 0.27 & 61.95 & 475.72 & 0.02 & 206.73 \\ \hline
\end{tabular}
\caption[Tabulated Individual Trial Statistics] {Mean and variance for the number of responses, the response rate, and peak times for the Balci and Church 1998 datasets.}
\label{Table_Balci2009_Individual}
\end{table}

\begin{table}[!ht]
\begin{tabular}{|lllllllll|}
\hline
Paper & Dataset & Simulation & \begin{tabular}[c]{@{}l@{}}Mean Num \\ Resp\end{tabular} & \begin{tabular}[c]{@{}l@{}}Mean Resp \\ Rate\end{tabular} & \begin{tabular}[c]{@{}l@{}}Mean Peak \\ Time\end{tabular} & \begin{tabular}[c]{@{}l@{}}Var Num \\ Resp\end{tabular} & \begin{tabular}[c]{@{}l@{}}Var Resp \\ Rate\end{tabular} & \begin{tabular}[c]{@{}l@{}}Var Peak \\ Time\end{tabular} \\ \hline
Church 2000 & PI15 Noise & PRDDM & 14.93 & 0.40 & 15.43 & 21.69 & 0.02 & 13.25 \\
Church 2000 & PI15 Noise & TDDM & 11.90 & 0.32 & 15.20 & 14.75 & 0.01 & 6.47 \\
Church 2000 & PI15 Noise & Data & 15.30 & 0.41 & 15.91 & 66.61 & 0.05 & 14.06 \\ \hline
Church 2000 & PI15 Light & PRDDM & 11.55 & 0.31 & 15.32 & 12.80 & 0.01 & 12.52 \\
Church 2000 & PI15 Light & TDDM & 9.80 & 0.26 & 15.58 & 9.55 & 0.01 & 4.47 \\
Church 2000 & PI15 Light & Data & 11.43 & 0.31 & 15.65 & 31.88 & 0.02 & 10.27 \\ \hline
Church 2000 & PI30 Noise & PRDDM & 20.54 & 0.27 & 31.68 & 27.10 & 0.00 & 35.75 \\
Church 2000 & PI30 Noise & TDDM & 18.51 & 0.25 & 30.07 & 39.00 & 0.01 & 17.38 \\
Church 2000 & PI30 Noise & Data & 20.62 & 0.27 & 32.24 & 88.63 & 0.02 & 23.29 \\ \hline
Church 2000 & PI30 Light & PRDDM & 22.14 & 0.30 & 30.37 & 35.25 & 0.01 & 47.14 \\
Church 2000 & PI30 Light & TDDM & 24.88 & 0.33 & 30.13 & 72.53 & 0.01 & 32.10 \\
Church 2000 & PI30 Light & Data & 22.44 & 0.30 & 30.84 & 75.01 & 0.01 & 44.12 \\ \hline
Church 2000 & PI60 90 Noise & PRDDM & 16.51 & 0.11 & 61.20 & 24.22 & 0.00 & 202.51 \\
Church 2000 & PI60 90 Noise & TDDM & 25.11 & 0.17 & 61.46 & 124.41 & 0.01 & 51.30 \\
Church 2000 & PI60 90 Noise & Data & 17.01 & 0.11 & 64.97 & 103.94 & 0.00 & 206.51 \\ \hline
Church 2000 & PI60 90 Light & PRDDM & 19.97 & 0.13 & 61.55 & 27.43 & 0.00 & 181.93 \\
Church 2000 & PI60 90 Light & TDDM & 24.96 & 0.17 & 61.48 & 119.13 & 0.01 & 45.15 \\
Church 2000 & PI60 90 Light & Data & 20.63 & 0.14 & 63.28 & 183.45 & 0.01 & 162.92 \\ \hline
Church 2000 & PI60 180 Noise & PRDDM & 18.18 & 0.12 & 60.52 & 23.42 & 0.00 & 155.10 \\
Church 2000 & PI60 180 Noise & TDDM & 35.38 & 0.24 & 59.36 & 149.72 & 0.01 & 65.48 \\
Church 2000 & PI60 180 Noise & Data & 15.84 & 0.11 & 61.93 & 61.53 & 0.00 & 101.18 \\ \hline
Church 2000 & PI60 180 Light & PRDDM & 18.46 & 0.12 & 59.93 & 26.33 & 0.00 & 204.09 \\
Church 2000 & PI60 180 Light & TDDM & 31.35 & 0.21 & 61.76 & 164.85 & 0.01 & 65.16 \\
Church 2000 & PI60 180 Light & Data & 17.19 & 0.11 & 61.83 & 78.66 & 0.00 & 207.04 \\ \hline
Church 2000 & PI120 Noise & PRDDM & 15.17 & 0.06 & 113.73 & 17.59 & 0.00 & 618.18 \\
Church 2000 & PI120 Noise & TDDM & 56.99 & 0.24 & 118.78 & 545.28 & 0.01 & 215.77 \\
Church 2000 & PI120 Noise & Data & 14.53 & 0.06 & 118.62 & 95.78 & 0.00 & 518.10 \\ \hline
Church 2000 & PI120 Light & PRDDM & 19.46 & 0.08 & 113.04 & 33.00 & 0.00 & 650.49 \\
Church 2000 & PI120 Light & TDDM & 65.86 & 0.27 & 119.12 & 802.19 & 0.01 & 340.04 \\
Church 2000 & PI120 Light & Data & 19.03 & 0.08 & 116.57 & 228.41 & 0.00 & 1045.29 \\ \hline
\end{tabular}
\caption[Tabulated Individual Trial Statistics] {Mean and variance for the number of responses, the response rate, and peak times for the Church 2000 dataset.}
\label{Table_Church2000_Individual}
\end{table}
\end{landscape}

\begin{landscape}
\begin{table}[!ht]
\begin{tabular}{|lllllllllll|}
\hline
Paper & Dataset & Simulation & Mean $S_1$ & Mean $S_2$ & Mean $M$ & Mean $D$  & Var $S_1$ & Var $S_2$ & Var $M$ & Var $D$ \\ \hline
Balci 2009 & PI30 & PRDDM & 20.21 & 47.79 & 34.00 & 27.58 & 54.99 & 45.32 & 39.83 & 41.30 \\
Balci 2009 & PI30 & TDDM & 12.54 & 52.15 & 32.34 & 39.61 & 76.53 & 41.86 & 38.29 & 83.60 \\
Balci 2009 & PI30 & Dataset & 13.88 & 44.84 & 29.36 & 30.97 & 77.75 & 93.27 & 64.11 & 85.58 \\ \hline
Church 1998 & PI30 & PRDDM & 14.70 & 46.50 & 30.60 & 31.81 & 36.05 & 59.35 & 35.34 & 49.45 \\
Church 1998 & PI30 & TDDM & 14.46 & 45.85 & 30.15 & 31.39 & 46.62 & 102.32 & 45.81 & 114.65 \\
Church 1998 & PI30 & Dataset & 15.76 & 46.20 & 30.98 & 30.45 & 62.39 & 78.11 & 51.10 & 76.58 \\ \hline
Church 1998 & PI45 & PRDDM & 25.05 & 64.40 & 44.72 & 39.34 & 100.05 & 116.53 & 84.59 & 94.78 \\
Church 1998 & PI45 & TDDM & 25.84 & 64.51 & 45.17 & 38.67 & 87.50 & 140.11 & 75.58 & 152.88 \\
Church 1998 & PI45 & Dataset & 27.09 & 63.83 & 45.46 & 36.75 & 109.12 & 117.31 & 85.74 & 109.90 \\ \hline
Church 1998 & PI60 & PRDDM & 34.60 & 89.45 & 62.03 & 54.85 & 161.85 & 188.80 & 141.80 & 134.08 \\
Church 1998 & PI60 & TDDM & 37.53 & 83.15 & 60.34 & 45.62 & 141.65 & 229.65 & 114.09 & 286.26 \\
Church 1998 & PI60 & Dataset & 37.68 & 85.73 & 61.71 & 48.05 & 188.70 & 253.35 & 162.22 & 235.22 \\ \hline
\end{tabular}
\caption[Tabulated Start and Stop Mean and Variance] {Mean and variance for the start ($S_1$), stop ($S_2$), mid-point ($M$), and duration ($D$) for the Balci 2009 and Church 1998 datasets.}
\label{Table_Balci2009_StartStop_MeanVar}
\end{table}

\begin{table}[!ht]
\begin{tabular}{|llllllll|}
\hline
Paper & Dataset & Simulation & $Corr(S_1,S_2)$ & $Corr(S_1,M)$ & $Corr(M,D)$ & $CV_{Start}$ & $CV_{Stop}$ \\ \hline
Balci 2009 & PI30 & PRDDM & 0.591 & -0.535 & -0.119 & 0.367 & 0.141 \\
Balci 2009 & PI30 & TDDM & 0.307 & -0.739 & -0.306 & 0.698 & 0.124 \\
Balci 2009 & PI30 & Dataset & 0.502 & -0.429 & 0.105 & 0.635 & 0.215 \\ \hline
Church 1998 & PI30 & PRDDM & 0.497 & -0.310 & 0.279 & 0.408 & 0.166 \\
Church 1998 & PI30 & TDDM & 0.248 & -0.403 & 0.384 & 0.472 & 0.221 \\
Church 1998 & PI30 & Dataset & 0.458 & -0.440 & 0.126 & 0.501 & 0.191 \\ \hline
Church 1998 & PI45 & PRDDM & 0.564 & -0.402 & 0.092 & 0.399 & 0.168 \\
Church 1998 & PI45 & TDDM & 0.337 & -0.433 & 0.245 & 0.362 & 0.184 \\
Church 1998 & PI45 & Dataset & 0.515 & -0.464 & 0.042 & 0.386 & 0.170 \\ \hline
Church 1998 & PI60 & PRDDM & 0.619 & -0.364 & 0.098 & 0.368 & 0.154 \\
Church 1998 & PI60 & TDDM & 0.236 & -0.492 & 0.243 & 0.317 & 0.182 \\
Church 1998 & PI60 & Dataset & 0.473 & -0.405 & 0.165 & 0.365 & 0.186 \\ \hline
\end{tabular}
\caption[Tabulated Start and Stop Correlations and Co-Variance] {Correlation and Co-Variance values for  start, stop, mid-point, and duration for the Balci 2009 and Church 1998 datasets.}
\label{Table_Balci2009_StartStop_Corr}
\end{table}

\begin{table}[!ht]
\begin{tabular}{|lllllllllll|}
\hline
Paper & Dataset & Simulation & Mean $S_1$ & Mean $S_2$ & Mean $M$ & Mean $D$  & Var $S_1$ & Var $S_2$ & Var $M$ & Var $D$ \\ \hline
Church 2000 & PI15 Noise & PRDDM & 8.16 & 22.01 & 15.08 & 13.85 & 10.82 & 16.22 & 9.99 & 14.10 \\
Church 2000 & PI15 Noise & TDDM & 9.76 & 19.71 & 14.73 & 9.95 & 7.58 & 11.08 & 5.69 & 14.59 \\
Church 2000 & PI15 Noise & Dataset & 9.39 & 22.35 & 15.87 & 12.96 & 11.34 & 15.66 & 8.29 & 20.84 \\ \hline
Church 2000 & PI15 Light & PRDDM & 8.37 & 21.96 & 15.16 & 13.59 & 9.67 & 13.59 & 7.41 & 16.86 \\
Church 2000 & PI15 Light & TDDM & 11.32 & 19.19 & 15.26 & 7.87 & 4.05 & 8.78 & 4.04 & 9.46 \\
Church 2000 & PI15 Light & Dataset & 10.32 & 21.40 & 15.86 & 11.08 & 7.43 & 14.41 & 7.41 & 14.04 \\ \hline
Church 2000 & PI30 Noise & PRDDM & 16.58 & 46.94 & 31.76 & 30.36 & 36.15 & 55.26 & 29.54 & 64.67 \\
Church 2000 & PI30 Noise & TDDM & 21.44 & 38.07 & 29.75 & 16.64 & 22.58 & 30.64 & 16.87 & 38.95 \\
Church 2000 & PI30 Noise & Dataset & 21.29 & 43.69 & 32.49 & 22.40 & 34.57 & 40.48 & 22.84 & 58.76 \\ \hline
Church 2000 & PI30 Light & PRDDM & 15.05 & 44.92 & 29.99 & 29.87 & 38.29 & 59.30 & 34.47 & 57.31 \\
Church 2000 & PI30 Light & TDDM & 18.12 & 41.14 & 29.63 & 23.02 & 34.26 & 62.37 & 30.35 & 71.85 \\
Church 2000 & PI30 Light & Dataset & 18.28 & 44.43 & 31.35 & 26.15 & 44.23 & 62.63 & 36.40 & 68.14 \\ \hline
Church 2000 & PI60 90 Noise & PRDDM & 32.37 & 91.43 & 61.90 & 59.06 & 188.95 & 267.56 & 159.60 & 274.63 \\
Church 2000 & PI60 90 Noise & TDDM & 49.21 & 73.32 & 61.27 & 24.11 & 51.81 & 111.01 & 49.27 & 128.56 \\
Church 2000 & PI60 90 Noise & Dataset & 46.62 & 84.97 & 65.79 & 38.34 & 116.55 & 221.25 & 97.96 & 283.75 \\ \hline
Church 2000 & PI60 90 Light & PRDDM & 31.73 & 92.92 & 62.33 & 61.19 & 159.37 & 250.63 & 140.43 & 258.26 \\
Church 2000 & PI60 90 Light & TDDM & 49.22 & 73.24 & 61.23 & 24.02 & 51.27 & 104.52 & 46.39 & 126.00 \\
Church 2000 & PI60 90 Light & Dataset & 42.87 & 88.99 & 65.93 & 46.12 & 108.61 & 304.86 & 116.69 & 360.17 \\ \hline
Church 2000 & PI60 180 Noise & PRDDM & 31.11 & 90.46 & 60.79 & 59.35 & 150.34 & 201.06 & 120.33 & 221.48 \\
Church 2000 & PI60 180 Noise & TDDM & 42.26 & 75.91 & 59.09 & 33.65 & 94.88 & 109.66 & 65.01 & 149.05 \\
Church 2000 & PI60 180 Noise & Dataset & 41.54 & 80.88 & 61.21 & 39.34 & 126.67 & 120.26 & 74.91 & 194.25 \\ \hline
Church 2000 & PI60 180 Light & PRDDM & 30.97 & 91.15 & 61.06 & 60.18 & 182.50 & 245.93 & 148.67 & 262.17 \\
Church 2000 & PI60 180 Light & TDDM & 46.46 & 76.71 & 61.58 & 30.25 & 65.77 & 148.28 & 64.83 & 168.78 \\
Church 2000 & PI60 180 Light & Dataset & 41.71 & 83.99 & 62.85 & 42.28 & 133.15 & 266.02 & 118.47 & 324.47 \\ \hline
Church 2000 & PI120 Noise & PRDDM & 68.49 & 165.52 & 117.00 & 97.03 & 678.29 & 689.57 & 466.20 & 870.94 \\
Church 2000 & PI120 Noise & TDDM & 90.57 & 146.42 & 118.50 & 55.85 & 310.18 & 390.86 & 214.93 & 542.32 \\
Church 2000 & PI120 Noise & Dataset & 91.11 & 155.32 & 123.22 & 64.21 & 554.26 & 553.20 & 330.70 & 892.12 \\ \hline
Church 2000 & PI120 Light & PRDDM & 64.48 & 166.63 & 115.56 & 102.15 & 647.77 & 746.35 & 498.18 & 795.52 \\
Church 2000 & PI120 Light & TDDM & 86.45 & 151.21 & 118.83 & 64.76 & 403.65 & 661.89 & 336.05 & 786.88 \\
Church 2000 & PI120 Light & Dataset & 87.80 & 159.24 & 123.52 & 71.44 & 678.09 & 688.60 & 356.35 & 1307.97 \\ \hline
\end{tabular}
\caption[Tabulated Start and Stop Mean and Variance] {Mean and variance for the start ($S_1$), stop ($S_2$), mid-point ($M$), and duration ($D$) for the Church 2000 datasets.}
\label{Table_Church2000_StartStop_MeanVar}
\end{table}
\end{landscape}

\begin{landscape}

\begin{table}[!ht]
\begin{tabular}{|llllllll|}
\hline
Paper & Dataset & Simulation & $Corr(S_1,S_2)$ & $Corr(S_1,M)$ & $Corr(M,D)$ & $CV_{Start}$ & $CV_{Stop}$ \\ \hline
Church 2000 & PI15 Noise & PRDDM & 0.488 & -0.352 & 0.227 & 0.403 & 0.183 \\
Church 2000 & PI15 Noise & TDDM & 0.222 & -0.527 & 0.192 & 0.282 & 0.169 \\
Church 2000 & PI15 Noise & Dataset & 0.231 & -0.537 & 0.164 & 0.359 & 0.177 \\ \hline
Church 2000 & PI15 Light & PRDDM & 0.279 & -0.507 & 0.175 & 0.372 & 0.168 \\
Church 2000 & PI15 Light & TDDM & 0.282 & -0.382 & 0.382 & 0.178 & 0.154 \\
Church 2000 & PI15 Light & Dataset & 0.377 & -0.346 & 0.342 & 0.264 & 0.177 \\ \hline
Church 2000 & PI30 Noise & PRDDM & 0.299 & -0.471 & 0.219 & 0.363 & 0.158 \\
Church 2000 & PI30 Noise & TDDM & 0.271 & -0.521 & 0.157 & 0.222 & 0.145 \\
Church 2000 & PI30 Noise & Dataset & 0.218 & -0.586 & 0.081 & 0.276 & 0.146 \\ \hline
Church 2000 & PI30 Light & PRDDM & 0.423 & -0.387 & 0.236 & 0.411 & 0.171 \\
Church 2000 & PI30 Light & TDDM & 0.268 & -0.441 & 0.301 & 0.323 & 0.192 \\
Church 2000 & PI30 Light & Dataset & 0.368 & -0.453 & 0.185 & 0.364 & 0.178 \\ \hline
Church 2000 & PI60 90 Noise & PRDDM & 0.404 & -0.430 & 0.188 & 0.425 & 0.179 \\
Church 2000 & PI60 90 Noise & TDDM & 0.226 & -0.425 & 0.372 & 0.146 & 0.144 \\
Church 2000 & PI60 90 Noise & Dataset & 0.168 & -0.492 & 0.314 & 0.232 & 0.175 \\ \hline
Church 2000 & PI60 90 Light & PRDDM & 0.380 & -0.412 & 0.240 & 0.398 & 0.170 \\
Church 2000 & PI60 90 Light & TDDM & 0.203 & -0.453 & 0.348 & 0.145 & 0.140 \\
Church 2000 & PI60 90 Light & Dataset & 0.146 & -0.414 & 0.479 & 0.243 & 0.196 \\ \hline
Church 2000 & PI60 180 Noise & PRDDM & 0.374 & -0.468 & 0.155 & 0.394 & 0.157 \\
Church 2000 & PI60 180 Noise & TDDM & 0.272 & -0.565 & 0.075 & 0.230 & 0.138 \\
Church 2000 & PI60 180 Noise & Dataset & 0.213 & -0.640 & -0.027 & 0.271 & 0.136 \\ \hline
Church 2000 & PI60 180 Light & PRDDM & 0.392 & -0.454 & 0.161 & 0.436 & 0.172 \\
Church 2000 & PI60 180 Light & TDDM & 0.229 & -0.409 & 0.394 & 0.175 & 0.159 \\
Church 2000 & PI60 180 Light & Dataset & 0.198 & -0.461 & 0.339 & 0.277 & 0.194 \\ \hline
Church 2000 & PI120 Noise & PRDDM & 0.363 & -0.559 & 0.009 & 0.380 & 0.159 \\
Church 2000 & PI120 Noise & TDDM & 0.228 & -0.563 & 0.118 & 0.194 & 0.135 \\
Church 2000 & PI120 Noise & Dataset & 0.194 & -0.635 & -0.001 & 0.258 & 0.151 \\ \hline
Church 2000 & PI120 Light & PRDDM & 0.430 & -0.485 & 0.078 & 0.395 & 0.164 \\
Church 2000 & PI120 Light & TDDM & 0.270 & -0.469 & 0.251 & 0.232 & 0.170 \\
Church 2000 & PI120 Light & Dataset & 0.043 & -0.689 & 0.008 & 0.297 & 0.165 \\ \hline
\end{tabular}
\caption[Tabulated Start and Stop Correlations and Co-Variance] {Correlation and Co-Variance values for  start, stop, mid-point, and duration for the Church 2000 datasets.}
\label{Table_Church2000_StartStop_Corr}
\end{table}
\end{landscape}

\end{appendices}

\bibliographystyle{apalike} 
\bibliography{skeleton}

\end{document}